\documentclass[fleqn,usenatbib]{mnras}
\usepackage{newtxtext,newtxmath}
\usepackage[T1]{fontenc}
\usepackage{ae,aecompl}
\usepackage{amssymb}
\usepackage{amsmath}
\usepackage{graphicx}
\usepackage{flushend}
\usepackage{microtype}
\def \Sw{{\it Swift}\ }

\def \upd{}
\def \NH{$N_\mathrm{H}$}

\title[The 2016 eruption of nova LMC\,1968]{The January 2016 eruption of recurrent nova LMC\,1968}

\author[Kuin et al.]{N.~P.~M. Kuin,$^1$\thanks{email: n.kuin@ucl.ac.uk}
K.~L. Page,$^2$
P. Mr\'oz,$^3$
M.~J. Darnley,$^4$
S. N. Shore,$^{5,6}$
\newauthor
J.~P. Osborne,$^2$
F. Walter,$^7$
F. Di Mille,$^8$
N. Morrell,$^8$
U. Munari,$^9$
T. Bohlsen,$^{10}$
\newauthor
A. Evans,$^{11}$  
R.~D. Gehrz,$^{12}$ 
S. Starrfield,$^{13}$  
M. Henze,$^{14,15}$        
S.~C. Williams,$^{16}$ 
\newauthor
G.~J. Schwarz,$^{17}$    
A. Udalski,$^3$ 
M.~K. Szyma\'nski,$^3$ 
R. Poleski,$^{3,18}$ 
I. Soszy\'nski,$^3$ 
\newauthor
V.~A.~R.~M. Ribeiro\,$^{19,20}$
R. Angeloni,$^{21,22}$ 
A.~A. Breeveld,$^1$
A.~P. Beardmore$^{2}$
\newauthor
J. Skowron,$^3$
\\
$^1$Mullard Space Science Laboratory, University College London, Holmbury St.\ Mary, Dorking, Surrey RH5 6NT, UK\\
$^2$X-Ray and Observational Astronomy Group, School of Physics \& Astronomy, University of Leicester, LE1 7RH, UK\\
$^3$Astronomical Observatory, University of Warsaw, Al. Ujazdowskie 4, 00-478 Warszawa, Poland\\
$^4$Astrophysics Research Institute, Liverpool John Moores University, IC2, Liverpool Science Park, Liverpool, L3 5RF, UK\\
$^5$Dipartimento di Fisica ``Enrico Fermi'', Universit\`{a} di Pisa, 56127 Pisa, Italy\\
$^{6}$NFN-Sezione Pisa, largo B.\,Pontecorvo 3, 56127 Pisa, Italy\\
$^{7}$Department of Physics and Astronomy, Stony Brook University, Stony Brook, NY 11794-3800, USA\\
$^{8}$Las Campanas Observatory, Carnegie Observatories, Casilla 601, La Serena, Chile\\
$^{9}$INAF-Astronomical Observatory of Padova, I-36012 Asiago (VI), Italy\\
$^{10}$Mirranook Observatory, Boorolong Rd Armidale, NSW, 2350, Australia\\
$^{11}$Astrophysics Group, Keele University, Keele, Staffordshire, ST5 5BG, UK\\
$^{12}$Minnesota Institute for Astrophysics, School of Physics \& Astronomy, University of Minnesota, Minneapolis, MN 55455, USA\\
$^{13}$School of Earth and Space Exploration, Arizona State University, Box 871404, Tempe, AZ 85287-1404, USA\\
$^{14}$Institut de Ci\`{e}ncies de l'Espai (CSIC-IEEC), Campus UAB, C/Can Magrans s/n, E-08193 Cerdanyola del Valles, Spain\\
$^{15}$Department of Astronomy, San Diego State University, San Diego, CA 92182, USA\\
$^{16}$Physics Department, Lancaster University, Lancaster, LA1 4YB, UK\\
$^{17}$American Astronomical Society, 2000 Florida Ave., NW, Suite 400, DC 20009-1231, USA\\
$^{18}$Department of Astronomy, Ohio State University, 140 W 18$^{th}$ Ave., Columbus, OH 43210, USA\\
$^{19}$ CIDMA, Departamento de F\'isica, Universidade de Aveiro, Campus Universit\'ario de Santiago, 3810-193 Aveiro, Portugal\\
$^{20}$ Instituto de Telecomunica\c{c}\~oes, Campus Universit\'ario de Santiago, 3810-193 Aveiro, Portugal\\
$^{21}$ Instituto de Investigaci\'on Multidisciplinar en Ciencia y Tecnolog\'ia, Universidad de La Serena, Av. R. Bitr\'an 1305, La Serena, Chile\\
$^{22}$ Departamento de F\'isica y Astronom\'ia, Universidad de La Serena, Av. J. Cisternas 1200, La Serena, Chile\\
}

\date{Accepted XXX. Received YYY; in original form ZZZ}

\pubyear{2019}

\begin{document}
\label{firstpage}
\pagerange{\pageref{firstpage}--\pageref{lastpage}}
\maketitle


\begin{abstract}


We present a comprehensive review of all observations of the eclipsing  {\upd  r}ecurrent Nova LMC 1968 in the Large Magellanic Cloud which was previously observed in eruption in 1968, 1990, 2002, 2010, and most recently in 2016. We derive a {\upd probable} recurrence time of $6.2 \pm 1.2$~years and provide the ephemerides of the eclipse. In the ultraviolet-optical-IR photometry the light curve {\upd shows high variability right from the first observation around two days after eruption. Therefore no colour changes can be substantiated. } {\upd Outburst} spectra from 2016 and 1990 are very similar and are dominated by H and He lines longward of 2000\AA. Interstellar reddening is found to be E(B-V) = $0.07\pm0.01$. The super soft X-ray luminosity is lower than the Eddington luminosity and the X-ray spectra suggest the mass of the WD is larger than 1.3 M$_\odot$.  Eclipses in the light curve suggest that the system is at high orbital inclination. On day four after the eruption a recombination wave was observed in Fe~II ultraviolet absorption lines. Narrow line components are seen {\upd after day 6} and explained as being due to reionisation of ejecta from a previous eruption. The UV spectrum varies with orbital phase, {\upd in particular} a component of the  He~II\,1640~\AA~ emission line, which leads us to propose that early-on the inner WD Roche lobe might be filled with a bound opaque medium prior to the re-formation of an accretion disk. Both {\upd this} medium and the ejecta can cause the delay in the appearance of the soft X-ray source. 
\end{abstract}


\setcounter{figure}{0}

\setcounter{table}{0}


\section{Introduction}



Nova eruptions are the result of the high-speed ejection of a turbulent mass  \citep{2016A&A...595A..28C,2018A&A...613A...8F}
due to a thermonuclear runaway (TNR) occurring within the surface layer of a mass-accreting white dwarf \citep[WD; see][for recent review articles]{2008clno.book.....B,2014ASPC..490.....W,starrfield2016}. \citet{2julian} provides a recent review of X-ray observations of novae.
In these close, semi-detached, binary systems the non-degenerate low-mass donor can be a main sequence star, a sub-giant, or a red giant \citep[see][for a summary]{2012ApJ...746...61D}.  
Novae that have been observed in eruption just once -- the majority of systems -- make up the group of classical novae (CNe), of which there are around 2000 known systems across the Milky Way and nearby galaxies.  
All novae are predicted to repeat \citep{1995ApJ...445..789P,2005ApJ...623..398Y}, and a small subset, the recurrent novae (RNe) have been observed in eruption more than once.  Observed recurrence periods range from $P_\mathrm{rec}\simeq1$\,year \citep[for M31N\,2008-12a]{2014A&A...563L...9D} up to 98\,years \citep[V2487\,Ophiuchi]{2009AJ....138.1230P} -- although the upper end must be a selection effect.

The short inter-eruption timescales of the RNe are believed to be due to a combination of a high mass WD and a high mass accretion rate \cite[]{starfield.88b}; the RNe contain many of the highest mass WDs known.  Within the Milky Way there are ten confirmed RNe, including the sub-class prototypes RS\,Ophiuchi \cite[$P_\mathrm{rec}\simeq20$\,years; with a red giant donor; see][for detailed reviews]{evans2008} and U\,Scorpii \cite[mean $P_\mathrm{rec} = 10.3$\,years; sub-giant donor; see, e.g.,][and references therein]{2015ApJ...811...32P}.  \citet{2010ApJS..187..275S} provided a detailed review of the known Galactic RNe; this and similar work indicated that the required high mass accretion rate is in most cases provided by mass loss from an evolved donor (sub- or red giant).  \citet{2014ApJ...788..164P} went on to estimate that the true RN population ($10\leq P_\mathrm{rec}\leq100$\, years; A.\ Pagnotta, private communication) of the Milky Way may be as high as $25\pm10\%$ of all Galactic novae.
In recent years some very rapid recurrent novae (RRNe) have been found with mean $P_\mathrm{rec}<10$\,years.  
The best studied is M31N\,2008-12a which has been detected in eruption every year since 2008 \citep[see][]{2014A&A...563L...9D,2015A&A...580A..45D,2016ApJ...833..149D,2014A&A...563L...8H,2015A&A...580A..46H,henze2018,2014ApJ...786...61T}, with a mean $P_\mathrm{rec}=0.99\pm0.02$\,years  {\upd\citep{DH2019}, which is surrounded by the super-remnant of thousands of earlier eruptions \citep{12a-NSR}.} Theoretical studies of RRNe point to the presence very high mass WDs with high mass accretion rates and low ejected mass \citep[e.g.,][]{starfield.88b,1995ApJ...445..789P,2005ApJ...623..398Y,2013ApJ...777..136W}. 
Therefore, in RRNe the ejecta become transparent on a much shorter time scale (than their CN counterparts), which opens the opportunity to study the evolution of the underlying system just after the eruption.

Extragalactic systems with their known distance often prove more suitable environments for the study of nova populations, and our near neighbour M31 is by far the best studied example \citep[see][for a recent review]{2018cosp...42E.760D}.  
M31 has an observed nova rate of $65^{+16}_{-15}$\,year$^{-1}$ \citep{2006MNRAS.369..257D} and over 1100 suspected novae have been discovered in that host alone \cite[see][and their on-line database\footnote{\url{http://www.mpe.mpg.de/~m31novae/opt/}}]{2007A&A...465..375P,2010AN....331..187P}.  \citet{2015ApJS..216...34S} compiled a catalogue of 16 M31 RNe and indicated that up to a third of M31 nova eruptions may be due to recurrent novae ($P_\mathrm{rec}\leq100$\,years).  
Using an independent approach, \citet{2016ApJ...817..143W} indicated that $30^{+13}_{-10}\%$ of M31 novae harboured red giant donors -- with an implication of a high mass accretion rate -- and that these systems were strongly associated with the disk of M31.  Although Nova LMC\,1968 (N\ LMC~1968), the subject of this paper, was the first {\it confirmed} extragalactic nova, (see Section 2), \citet{2015ApJS..216...34S} reported that the M31 nova M31N\,1926-06a \citep{1929ApJ....69..103H} was the first extragalactic nova to be observed to recur \cite[as M31N\,1962-11a; see][]{1964AnAp...27..498R,1968AN....291...19B,2008A&A...477...67H}

The first nova in the Large Magellanic Cloud (LMC; LMCN\,1926-09a) was reported by \citet{1927BHarO.847....8L}.  Since then there have been {\upd 50 unique } LMC nova candidates \citep[see][and on-line catalogue$^1$]{2013AJ....145..117S}, 
around half of which have been spectroscopically confirmed.  \citet{2016ApJS..222....9M} recently computed the global nova rate of the LMC to be $2.4\pm0.8$\,year$^{-1}$. Within the LMC there are four known recurrent novae, YY\,Doradus \citep[LMCN 1937;][]{2004IAUC.8424....1B,2004IAUC.8424....2M}, LMCN\,1971-08a \citep{2016ApJ...818..145B}, OGLE-2018-NOVA-01 \citep[LMCN 2018-02a;][]{mroz2018a} and Nova LMC\,1968 
\citep[]{shore}.  \citet{2013AJ....145..117S} concluded that $\sim10\%$ of LMC novae are recurrent, and that $\sim16\%$ of observed LMC eruptions occur from RN systems, though uncertainties are large since these numbers are based on four RNe only. 
For comparison, 22 nova candidates have been discovered in the Small Magellanic Cloud$^1$  
which with its known distance provides good multi-spectral coverage opportunities  \citep{2018MNRAS.474.2679A}.  
\citet{2016ApJS..222....9M} compute a nova rate of $0.9\pm0.4$\,year$^{-1}$. There are no known RNe in the SMC.



The geometry of novae ejecta is non-spherical. Axially symmetric geometries were first discussed  by \citet[]{CecilaPG}. Morpho-kinematical emission line profile modelling with  SHAPE\footnote{http://bufadora.astrosen.unam.mx/shape/index.html} \citep{ribeiro2013,2013ApJ...768...49R} shows bipolar geometries. Recent Monte-Carlo based radiative transfer models of ejecta far enough after the eruption to ensure low densities and frozen-in state provide good fits to cone-shape geometries \citep{shore2013_tpyx,shore2013_mon,shore2016_v339del} and can include embedded dust \citep{2018A&A...619A.104S}. 

If a carbon-oxygen (CO) WD approaches and passes the \citet{1931ApJ....74...81C} mass a type Ia supernovae (SN\,Ia) thermonuclear explosion may ensue \citep[see, e.g.,][]{1973ApJ...186.1007W,2000ARA&A..38..191H, 2018PhR...736....1L}, whereas accreting oxygen-neon WDs are predicted to collapse to neutron stars \citep[see, e.g.,][]{1996ApJ...459..701G}.  
Novae have long been a proposed SN\,Ia progenitor pathway; the two difficulties have been the unknown  composition of the massive WDs found in RNe, and the relatively low number of systems.  However, recent work by \citet{2016ApJ...819..168H} has indicated that an accreting CO WD can grow from its formation mass $(<1.1\,\mathrm{M}_\odot)$ up to the Chandrasekhar mass with little or no tuning of the system parameters \citep[see also][]{2008NewAR..52..386H,starrfield2012}, whereas work by \citet{2014ApJ...788..164P,2015ApJS..216...34S,2016ApJ...817..143W} has indicated that the underlying size of the RN population may be larger than first determined, as may be the global nova rates themselves \citep[also see][]{2016MNRAS.458.2916C,2016ApJS..227....1S,2016MNRAS.455..668S,2017ApJ...834..196S}.

In this paper we present panchromatic observations, from the near infrared to the X-ray, of the 2016 Jan.\ eruption of the rapidly recurring N\  LMC\,1968. We also revisit observations from previous eruptions. In Section~\ref{1968} we provide a summary of the eruption history of the system. 
In Section~\ref{sec:observations} we present the observations, give the orbital ephemerides, the light curve and X-ray data analyses. 
In Section~\ref{sec:obs_results} the WD mass, reddening, secondary star, the discovery of a  recombination wave in Fe$^+$, and narrow line profiles are discussed. 
In Section~\ref{sec:model} an estimate of the mass ejected, He mass addition from H burned during the SSS phase, orbital inclination, and accretion disk formation are discussed in the context of a comprehensive model. 
Section~\ref{sec:results} provides a brief summary of some of the main points and a table summarizing the derived parameters.

\section{The Recurrent Nova LMC~1968}
\label{1968}


N\ LMC\,1968 \citep[originally referred to as Nova Mensae\,1968;][]{sievers} was first discovered in eruption at a magnitude of $m_\mathrm{pg}=10.9$ on Bamberg plates taken by I.~Paterson on 1968 Dec.\ 16.5~UT.  
By virtue of its decline rate of 0.5\,mag\,d$^{-1}$ (see Section \ref{ogle}), the speed class \citep[see][]{CecilaPG} of this eruption was classified as `fast'.  
The best estimate for the onset of the 1968 eruption is MJD\,$40206\pm1$ \citep{sievers}. Subsequently, this nova has been seen in eruption a further four times, most recently in 2016 Jan.  The 1990, 2002, and 2010 eruptions are summarised below, and the 2016 eruption is introduced.

\begin{figure}
\includegraphics[width=\columnwidth]{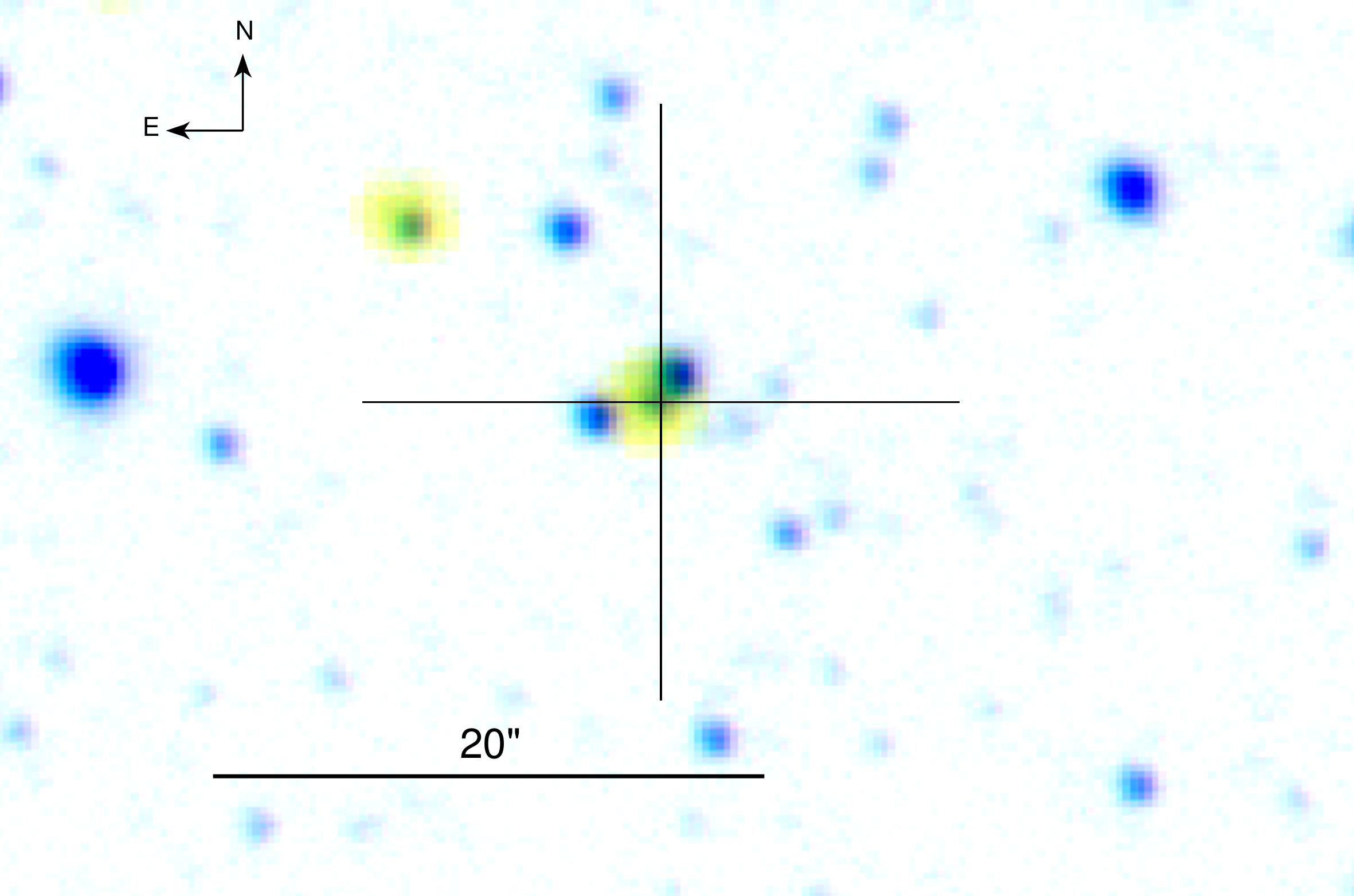}
\caption{The finding chart of the nova.  North is up and East is to the left. 
RGB inverted composite of UVOT $uvm2$ (yellow), OGLE $I$-band (magenta), $V$-band (cyan). The black cross 
marks the nova position.}
\label{fig_ogle_finder}
\end{figure}
\subsection{Nova LMC 1990b}
\label{sec:nova1990b}
\begin{figure}
\includegraphics[width=\columnwidth]{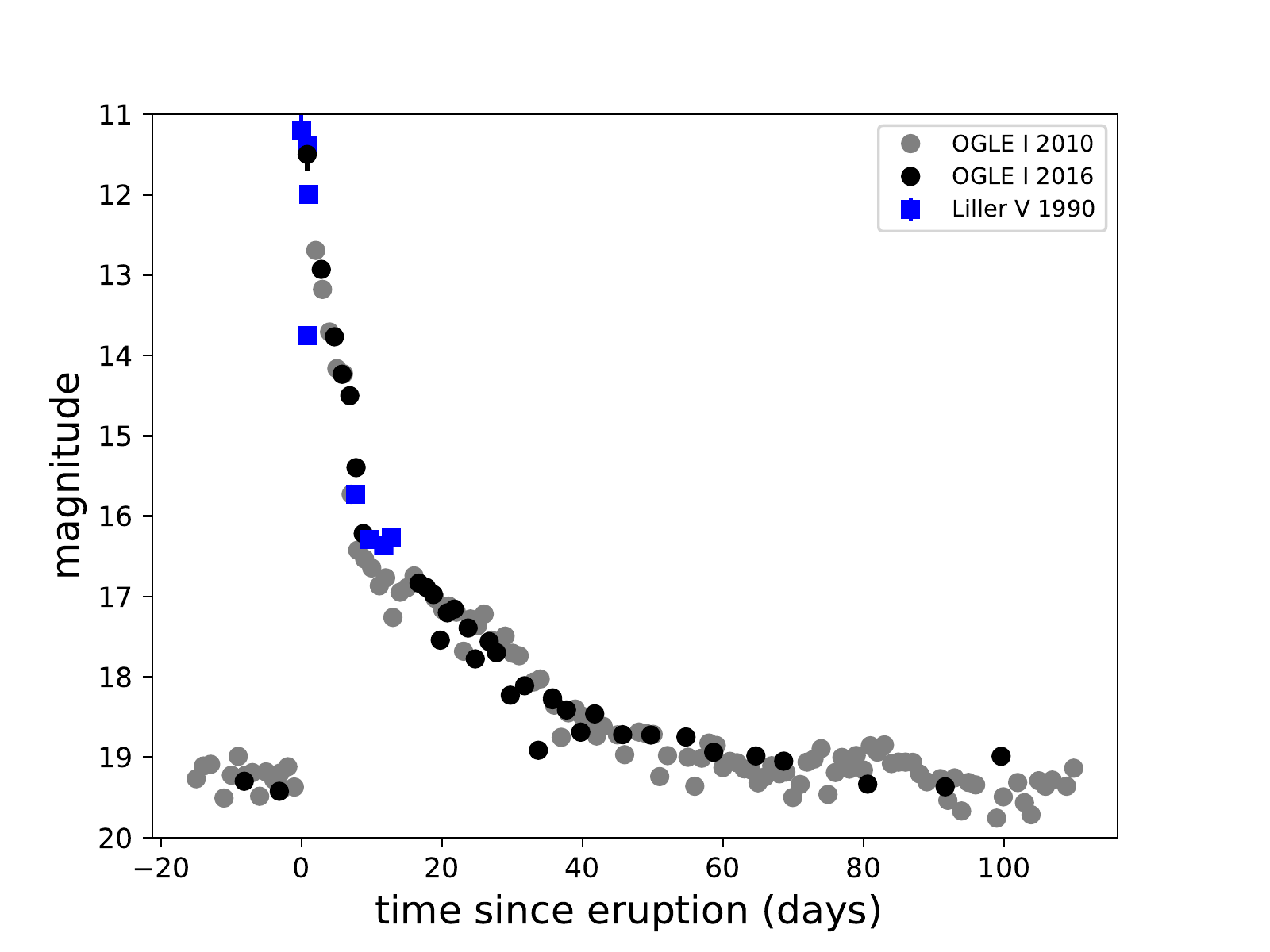}
\caption{A comparison between the OGLE $I$-band data from the 2010 and 2016 eruptions of N\ LMC\,1968. The V-band photometry from \protect\cite{liller2004} of the 1990 eruption is included as blue squares.}
\label{fig_ogle_lc}
\end{figure}

On 1990 Feb.\ 14.1~UT (MJD\,47936.1) Nova LMC\,1990b was discovered  in eruption at $m_\mathrm{pv} = 11.2$\,mag, close to the position of N\  LMC\,1968 \citep{liller}, see Fig.\ref{fig_ogle_finder} for a recent finder chart. 
A direct comparison of the Nova LMC\,1990b position to that of N\ LMC\,1968 using the original photographic plates showed a match in R.A.\ within $2\farcs4$ and in Dec.\ within 6\arcsec~ \citep{shore}. 
The peak time of eruption is more difficult to establish, as the last pre-eruption observation 
was on 1990 Feb.\ 3 \citep{liller2004}, 11 days pre-discovery. However, the photometry matches that of later eruptions well (see Fig. \ref{fig_ogle_lc}) and based on that the discovery time would be  within 0.2\,d of the eruption.  

Optical spectra of the 1990 eruption were obtained by Shara and Moffat 
\citep{williams} providing confirmation of the nova, thereby identifying Nova LMC\,1990b as the first {\it spectroscopically confirmed} extragalactic RN and on day 9 by \citet{williams} at the Cerro Tololo Inter-American Observatory (CTIO) in Chile. 
\citet{sekiguchi} obtained 
photometry and spectra on day 8.7, 9.7, 11.7, and 15.7 at the SAAO in Sutherland, South Africa.  
Judging by the comparison of the 1990 magnitude at discovery to the well-observed 2010 and 2016 light curves, the (now lost) discovery spectrum was likely obtained within a day post-eruption though no time was reported in the IAU Circular 4964 of 15 Feb. 1990. 
The discovery spectrum showed \ion{He}{i} and \ion{He}{ii} 4686\,\AA\  lines with expansion velocities of 5500\,km\,s$^{-1}$, and broad double-peaked Balmer lines \citep{williams}, 
which ``one tends not to see that shortly after eruption'' (R.~E. Williams, 
private communication).  The 1990 eruption was deemed to be spectrally similar to the Galactic recurrent nova U\,Sco, with a similarly fast evolution.

Nova LMC\,1990b was well observed with the International Ultraviolet Explorer (IUE) 
satellite which covered the eruption in the 1050--3250\,\AA\ wavelength range 
starting just 2 days after the discovery of the 1990 eruption, 
and observing on days 3, 5, 9, 14, 32, and 38 \citep{shore}.  
The He/H ratio derived from the IUE spectra by \citet{shore} was exceptionally high and seen as evidence for an evolved companion. 
The total UV luminosity was shown by \citet{shore} to be large, possibly exceeding  the Eddington luminosity for a Solar mass WD. 
\citet{shore} used a value of $E(B-V)=0.15$\,mag and the Fitzpatrick extinction curve \citep{fitzpatrick1986} and thus applied an extinction correction larger than we currently believe is correct (see \ref{sec:reddening}).
Since 1990, the LMC distance has also been revised downward from 55\,kpc to $49.59\pm0.60$\,kpc \citep{2013Natur.495...76P, 2017AJ....153..154B,Pietrzy2019}, which is the distance we adopt in this paper. 
We revisit below the question of the luminosity in light of the reduced extinction and distance. 
\citet{shore} derived an accretion rate M$_{acc} \ge 10^{-8.6 \pm 0.5} {\rm M}_{\sun} {\rm yr}^{-1}$ based on an estimated mass ejection of M$_{ej}$ = 10$^{-7.3\pm 0.5} {\rm M}_{\sun}$ which may have been overestimated; the mean velocity width of the UV lines (FWZI) was 12,000 km s$^{-1}$ with a FWHM of 5000 km s$^{-1}$.
The paucity of optical spectra of the 1990 eruption left open the question of 
whether the UV-optical derived He/H ratio was probing the same region of ejecta as 
the optical observations.

\subsection{Missed eruptions?}

The nova was almost continuously observed by the MACHO microlensing survey
\citep{1992ASPC...34..193A,alcock} during the years 1993--1999 and there 
is no evidence for further eruptions in the available data. 
Eruptions of N\ LMC\,1968 last $\sim50$\,days; however the initial 
decline essentially takes 15\,days. 
There are a few gaps in the MACHO data as long as 40--60 days and 
there is thus a small probability that the nova erupted during these
gaps.  
The limit for the MACHO observations in the LMC is V$ \approx$ 18 
\citep[the LMC sky background is R$\approx$19.5 mag/arcsec$^2$;][]{1999PASP..111.1539A}, so an eruption would be detected
significantly for up to 40 days, see Fig.\,\ref{fig_ogle_lc}.
Based on the MACHO coverage of the LMC and assuming the nova can no longer be observed 40 days after eruption, the probability for a missed eruption is just 4\%.  

\begin{table*}
\begin{minipage}{145mm}
\center
\caption{Known eruptions of N LMC 1968.\label{table_eruptions}}
\begin{tabular}{@{}llllllrrcc}
\hline
Discovery & \multicolumn{2}{c}{Positions$^1$}  &  adopted eruption&  discovery  &   \\ 
date     & \multicolumn{2}{c}{ RA, Dec (J2000)} & dates (MJD)                     & magnitude$^2$ &   \\
\hline
1968 Dec. 16.5& 05:10:00[06]    & -71:39:05[60]  &  40206.0 $\pm$ 1.5  &  m$_\mathrm{pg}=$10.9      & \\
1990 Feb. 14.1& 05:09:58.3[.5]  & -71:39:51.3[6] &  47936.1 $\pm$ ?    &  m$_\mathrm{pv}$=11.2      & \\
2002 Oct. 10& 05:09:59.4      & -71:39:51.5    &  52557.3 $\pm$ 1.0  &  V$ = 11.15 \pm 0.02$     & \\
2010 Nov. 21.2& 05:09:58.39[.01]& -71:39:52.7[.1] &  55521.2 $\pm$ 1.0  &  I$ = 11.7 \pm 0.3$ & \\
2016 Jan 21.2& 05:09:58.39[.01]& -71:39:52.7[.1] &  57407.4 $\pm$ 0.8  & I$ \approx 11.5 \pm 0.2$     & \\
\hline
\end{tabular}
\end{minipage}
\center{1 -- Positional uncertainties are given in square brackets. \\  2 -- The discovery magnitude is not neccesarily the peak magnitude.}
\end{table*}

\subsection{The 2002 eruption}

In 2002 the All Sky Automated Survey (ASAS-3) observed the nova region on every other day  
\citep{pojmanski}.  
On 2002 Oct. 9, 7:18 UT (MJD\,52556.304) the nova was not seen, but there were positive detections of the 2002 eruption in two subsequent observations on 2002 Oct. 11, 6:08 UT with V = $11.15 \pm 0.02$ and with V = $14.17 \pm 0.13$ on 2002 Oct. 15, 5:20 UT. 
The 2002 eruption was also observed by the Exp{\'e}rience pour la Recherche d'Objets Sombres (EROS) group 
\citep[P.~Tisserand \& J.-B. Marquette, private communication]{tisserand}. 
Unfortunately, a useful light curve could not be extracted from the EROS data. 
Based on the above we conclude that the 2002 eruption peaked  between MJD\,52556.3 and 52558.3 (i.e., 2002 Oct.\ 10).

\subsection{Nova LMC 2010}

The next detected eruption of N\ LMC\,1968 occurred on 2010 Nov.\ 21.2~UT 
(MJD\,55521.2$\pm$1.0) with a peak observed magnitude $11.7\pm0.3$ \citep{mroz}
and was discovered by the Optical Gravitational Lensing Experiment \citep[OGLE; see][]{2008AcA....58...69U,udalski}. 
The coordinates of the nova were measured to be $\alpha = 5^\mathrm{h}9^\mathrm{m}58\fs39$, $\delta = -71^\circ39^\prime52\farcs7$ (J2000) from  the OGLE data, with an astrometric accuracy of $0\farcs1$.  
A detailed analysis is provided in Section \ref{ogle}.

\subsection{Nova LMC 2016}
\label{sec:nova2016}
The most recent eruption of N\ LMC\,1968 was discovered by OGLE on 2016 Jan.\ 21.2094~UT at a reported  magnitude $I \le 11.5$ \citep{atel8578}.  The previous OGLE observation on Jan.\ 17.2363 indicated that the nova was still at quiescence ($I\simeq19$ mag).  
This eruption was assigned the internal OGLE designation OGLE-2016-NOVA-01.

The 2016 eruption must have peaked between Jan.\ 19.65000~UT 
(which is the last non-detection from the AAVSO\footnote{\url{https://www.aavso.org}} database) and Jan.\ 
21.20942~UT (the OGLE detection). 
The mean time is JD\,2457407.9 = Jan. 20.4 $\pm$ 0.8 
which we define as $t_0$, the best estimate of the time of eruption, see Table~\ref{table_eruptions}.
For the 2016 eruption on JD=2457408.709 we estimate the  magnitude, near the peak, is $I = 11.5\pm0.2$\,mag, close to those of previous eruptions.

\subsection{A rapidly recurring nova}
\label{recurrencetime}
Given the continuing monitoring from ASAS, and OGLE, the last three detected eruptions, 2002, 2010, 
and 2016 were likely to be subsequent eruptions.
However, it is possible that between 1968 and 1990 several eruptions were missed. 
Considering that we had inter-eruption intervals (going back in time) of 1887\,d, 2963\,d, 4621\,d, and 7731\,d, from the three most recent eruptions we get an estimate of $2425 \pm 540$\,d.
Using the first three intervals while  accounting for a missed eruption in 1996 gives a mean period of 2356\,d, which suggests either 2 or 3 missed eruptions between 1968 and 1990.  
If we assume there were 2 missed eruptions we get a  best estimate for the interval between eruptions of  $P_\mathrm{rec}=6.2\pm1.2$\,years, (2 standard deviations) with the possibly missed eruptions centred around February 1976  and October 1982.  

With $P_\mathrm{rec}<10$\,years, we can therefore consider N\ LMC\,1968 as one of the most rapid recurrents, and comparable to the Galactic RN U~Sco. 
With all the other extragalactic examples being in M31, N\ LMC\,1968 is the closest RRN with accurate  distance, and therefore should be studied in much greater detail.


\begin{figure*}
\includegraphics[width=162 mm]{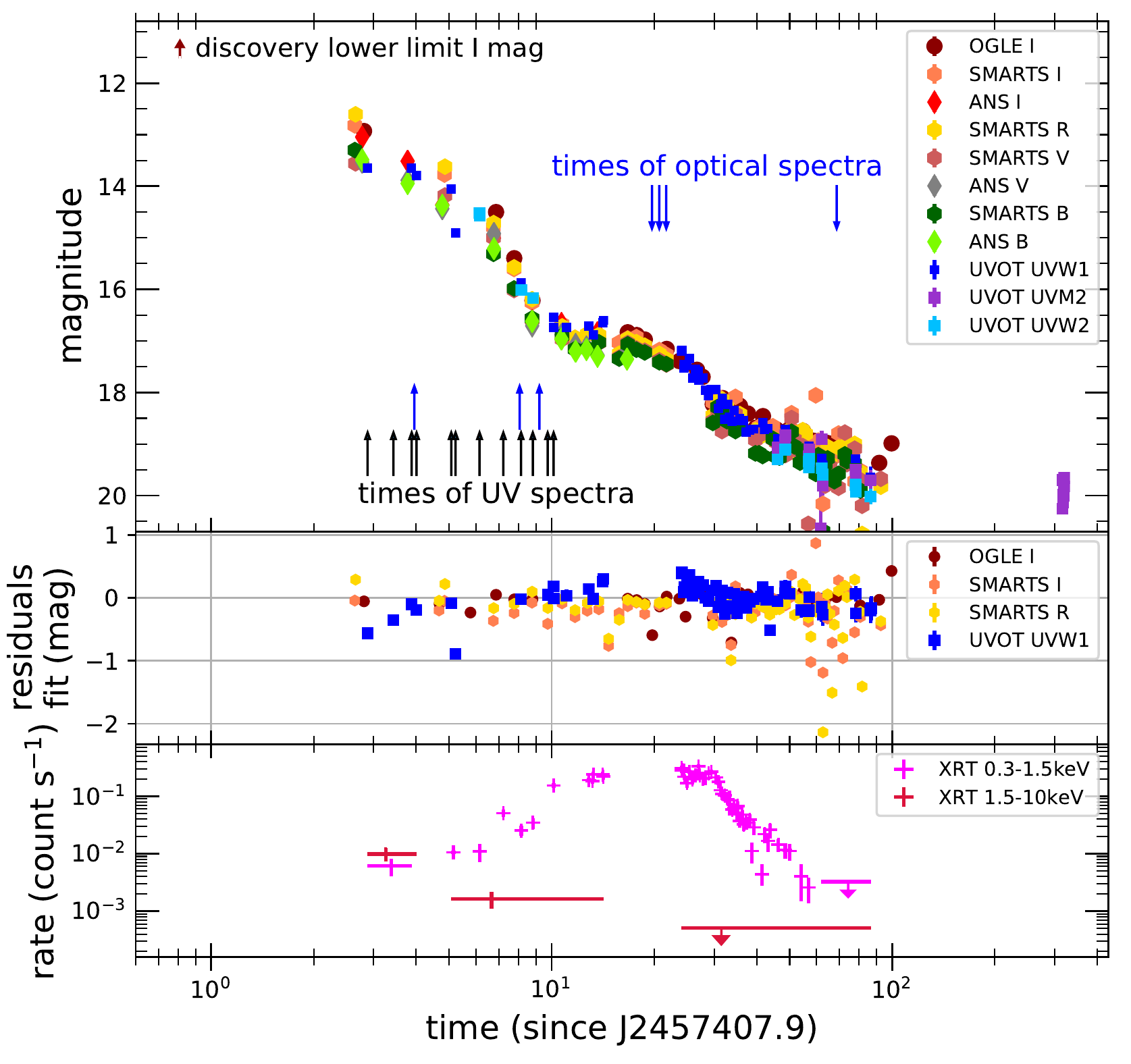}
\caption{The optical, UV and X-ray light curves for the 2016 eruption of Nova LMC\,1968. Error bars are included, but in the UV/optical are mostly smaller than the symbols. The first I-band was saturated, leading to a lower limit.
In the top panel data which fell within 0.1 of orbital phase 0.0 were excluded. 
The middle panel shows the residuals for $I$, $R$, and $uvw1$ from the light curve fit, see Table~\ref{tab:table_lc_slope}; these include all observations. 
All panels have a common time axis where times are from the estimated eruption time. 
The times when spectra were obtained have been indicated. 
Note that the $BVRI$ magnitudes are on the Vega system but the UV magnitudes are on the AB system. 
The X-ray light curves for the harder and softer photons have been separated to show the initial decline in hard photons and rise in soft emission.}
\label{fig_uv_opt_ir_lc}
\end{figure*}

%
\section{Multispectral Observations}

\label{sec:observations}

\subsection{\textit{Swift} observations (UVOT and XRT)}

Upon discovery of the N\ LMC\,2016 a Neil Gehrels \Sw Observatory \citep{swift} target of opportunity observation  started 
on 2016 Jan.\ 23~06:40 UT (day 2.88) with the X-ray Telescope 
\citep[XRT;][]{burrows} and the Ultraviolet and Optical Telescope
\citep[UVOT;][]{mason, roming} obtaining  photometric data in the UV filters 
and UV grism spectra. 
Regular \Sw observations continued until Apr.\ 16 (day 85.5). The final
XRT detection was on Mar.\ 17 (day 56.1), with the remaining 5.5\,ks of 
exposure only providing an X-ray upper limit.  Due to an observing 
constraint\footnote{There is an area near its orbital pole which is 
unobservable to {\em Swift}, due to the requirement that the spacecraft 
point more than $30\degr$ from the Earth limb, which subtends an angle of $\sim66\degr$ to 
the orbital altitude. This pole constraint can prevent observations of a 
given target for several days.}, there were no \Sw data between day\,14.1 and 
day\,24.1.
\Sw observations were not continued beyond this point  since by that 
time the XRT count rate was too low and the contamination of the photometry by nearby sources was thought to be too large.
However, additional UVOT observations were obtained on day 215--320.
The processing of the \Sw data is discussed in relation to similar multispectral 
data in the following sections.

\subsection{XRT data reduction, the soft-X-ray light curve \& variability}
\label{sec:xrtlightcurve}

The \Sw XRT observations started with a 500\,s observation  in Photon Counting (PC) mode.
The XRT detected faint hard X-ray emission at day 4 \citep{atel8587}. 
A significant increase in X-ray counts below 1 keV was detected on
Jan. 27, only six days after discovery. This indicated the emergence of the
supersoft X-ray source (SSS) emission - a phase that had never been
observed before for this nova.\footnote{We checked for pre-Swift observations of N LMC 1968 in  other X-ray and EUV mission catalogues; it was not  detected.}

The XRT data were processed and analysed using the standard 
{\sc heasoft} tools and most up-to-date calibration files. 
All the X-ray data were collected in PC mode, and grades 0--12 
were analysed for both the light-curves and spectra. No pile-up was evident 
at any time, so a circular region was used to extract the source counts, with 
the radius changing from 10 to 15\,pixels ($1\,\mathrm{pixel}\equiv2\farcs36$) depending on 
the brightness of the source. 

Figure~\ref{fig_uv_opt_ir_lc}  plots the soft (0.3--1.5\,keV) and hard 
(1.5--10\,keV) band light-curves. There are very few counts above 1.5\,keV, 
and this flux fades rapidly. The X-ray count rate was 
approximately constant from the day of first detection (day 2.9) until 
day 6.1; the soft emission then increased in brightness from day 7.2 until 
day 14.1, reaching $\sim0.25$\,count\,s$^{-1}$. At this point, the source 
became unobservable to \Sw. 
The nova re-emerged from the observing constraint on day 24.1 
at about the same count rate as ten days previously and, after a short 
plateau phase, started to fade from day 31. The X-ray source was no longer 
detected after day 57, with a final $3\sigma$ upper 
limit $3.3\times10^{-3}$\,count\,s$^{-1}$ (0.3--1.5\,keV) using 
5.5\,ks of data collected between 62 and 87 days. 

High-amplitude X-ray variability has been observed in several novae during the early SSS phase which is yet to be fully understood \citep[see, e.g.,][]{2009AJ....137.4160N,osborne2011,2016ApJ...818..145B}. 
These variations are seen to occur on time scales from hours to days. 
Despite daily coverage during the first 12 days of the eruption, no {large} amplitude X-ray variability was seen during the rise of the SSS phase in N\ LMC 1968.

\subsection{OGLE observations}


\label{ogle}
N\ LMC\,1968 has been monitored in the $V$- and $I$-bands 
as part of the ongoing OGLE survey since 2010. These observations were conducted using the 1.3\,m Warsaw Telescope
located at the Las Campanas Observatory, Chile. All data were reduced and calibrated 
following the standard OGLE pipeline \citep{udalski}.
Both the 2010 and 2016 eruptions were discovered by OGLE; the 2010 in archival data \citep{mroz,atel8578}.  

The nova has a close ($\approx1\farcs2$ distant) yet unrelated, bright, non-variable on-sky neighbour. 
Both stars are resolved in the OGLE photometry, see Figure~\ref{fig_ogle_finder} which is a  
colour-coded composite image that also includes the UVOT $uvm2$ band, which has a 
central wavelength of 2246\,\AA\ (and has a broader point
spread function than the ground-based data).  
OGLE uses differential image techniques, so the flux is measured on the 
subtracted images and the effects of blending due to neighbouring stars is reduced.

Because both the 2010 and 2016 eruptions have been observed by OGLE, we can compare data taken by the same instrument for two subsequent eruptions. 
In Figure~\ref{fig_ogle_lc} the $I$-band light curves of the 2010 and 2016 eruptions are shown for comparison, the similarity is clearly evident.  \citet{2010ApJS..187..275S} noted that all eruptions of a given RN appear essentially identical.

Using the estimated time of eruption,  we find a decrease in the $I$ band of two magnitudes after $t_2=3.9\pm0.8$\,d, and three magnitudes after $t_3=5.9\pm0.8$\,d, see Fig.~\ref{fig_uv_opt_ir_lc}.

\subsection{SMARTS/Andicam photometry}

\label{section_andicam}

$BVR_CI_C$ photometry (Vega magnitudes) was obtained with the Andicam 
instrument on the Small and Moderate Aperture Research Telescope System (SMARTS) 1.3\,m telescope at CTIO
starting in Feb. 2012, monitoring N\ LMC\,1968 twice a year \citep{smartsnovaatlas}. 
Daily observations of the 2016 eruption were conducted from day~2 
to 81 with a break between day 22--29, further observations on day 92 and 203 and continuing. No new eruption has been seen as of $t_0+1174$ days. 

As noted before, there are two stars in close proximity to the nova on the sky. 
To separate these stars, we fit three two-dimensional Gaussians 
to each observation using the {\tt IDL} routine {\tt MPFITFUN} \citep{markward}\footnote{
\url{http://purl.com/net/mpfit}}
constrained to have the same widths. 
The distribution in flux between the three
stars is given by the ratios of the best fit amplitudes. Since the data
are noisy, and we do not a priori know the positions of the stars,
we allow the centres of the Gaussians to wander within 1 pixel ($0\farcs37$)
of the median position.

We determined the relative positions from the fits on 14 nights, Jan.\ 29 through Feb.\ 22, excluding nights with particularly bad seeing.
The contribution of the nova to the summed flux dropped from 71\% to 24\%
during this time.
Relative to the nova, the mean offset positions are:
SE star: $\Delta\alpha = -2\farcs17 \pm 0\farcs06$, $\Delta\delta = -0\farcs34 \pm 0\farcs04$;
NW star: $\Delta\alpha= +0\farcs88 \pm 0\farcs07$, $\Delta\delta = +1\farcs08 \pm 0\farcs08$;
see Figure~\ref{fig_ogle_finder}.

We calibrated the fluxes using aperture photometry in an 11 pixel ($4\farcs06$)
aperture. We use 25 stars in the field as comparisons. They are calibrated
against Landolt standard stars on photometric nights to determine their
magnitudes.

\subsection {ANS photometry}

$B$$V$$I_{\rm C}$ optical photometry of the nova was obtained with the Asagio Novae and Symbiotic stars (ANS) collaboration \citep{2012BaltA..21...13M}  robotic telescope, described by \citet{2012BaltA..21...22M}, located in San Pedro de Atacama
(Chile).  
Detailed analysis of the photometric performances and multi-epoch measurements of the actual transmission profiles for the photometric filter sets in use is presented by \citet{2012BaltA..21...22M}.  Data reduction used the APASS sources for calibration \citep{2012JAVSO..40..430H,2014CoSka..43..174M}  using the transformation equation calibrated in \citet{2014JAD....20....4M,2014AJ....148...81M}.  The APASS survey is strictly linked to the \cite{2009AJ....137.4186L} and \cite{2002AJ....123.2121S} systems of equatorial standards.  

All measurements were carried out with aperture photometry, with the 
aperture radius and inner/outer radii for the sky annulus $\chi^2$-optimized 
on each image  to reduce dispersion of the stars in the local photometric 
sequences around the transformation equations from the local instantaneous to the standard system.  
Finally, colours and magnitudes were obtained separately during the reduction 
process, and were not derived one from the other. 
The quoted uncertainties include all error sources. 
Magnitudes are on the Vega system.
The nearby stars contaminate the photometry due to the large ANS PSF so that the data fainter than 16th magnitude cannot be used without a correction. There are enough faint data points to be useful.  
To fit ANS to the OGLE and SMARTS photometry the ANS magnitude was transformed 
using 
\begin{equation}
\mathrm{mag}_\mathrm{corr} = 2.5 \times \log_{10}\left[10^{-0.4\left(V+a\right)}-b\right]
\end{equation}
\noindent
with $a_I = 0.7, b_I = 8\times 10^{4}, a_V = 0.3, b_V = 3\times 10^{-8}, a_B = 0.4,$ 
and $b_B = 3\times 10^4$.

\citet{munari} report the early $BVRI$ photometry from 
SMARTS/Andicam and the Asagio Novae and ANS combined. 

\begin{table*}
\caption{Optical-UV light curve fitted parameters 
\label{tab:table_lc_slope}}
\begin{tabular}{@{}llllllllllllcr}
\hline
band   &start &\multicolumn{3}{c}{$t_\mathrm{break}$\ times} &     & \multicolumn{4}{c}{slope see (1)} && end & \\ 
       &time  &$t_a$&$t_b$&$t_c$&$t_d$& \multicolumn{4}{c}{}  &&time &\\
       & (d)  &(d)&(d)&(d)&(d)           & $s_a$ & $s_b$ & $s_c$ & $s_d$ & $s_e$ & (d) & \\ %
\hline
$I, I_c\,^3$   &0.81& 
$6.2\pm0.5$&
$9.3\pm0.4$&
$21.2\pm1.9$&
$32.0\pm3.4$&
$3.6\pm0.4$&
$16.1\pm0.4$&
$0.9\pm0.5$&
$5.1\pm0.3$&
$2.1\pm0.4$&
100&\\ 
$R$   &2.65& 
$5.8\pm0.3$&
$9.5\pm0.3$&
$19.9\pm1.9$&
$34.7\pm7.3$&
$3.9\pm0.4$&
$13.1\pm1.5$&
$1.3\pm0.6$&
$5.8\pm1.4$&
$2.5\pm0.3$
&93&\\
$V$   &2.65&
$6.1\pm0.3$&
$9.2\pm0.3$&
$19.6\pm1.7$&
$29.7\pm9.6$&
$2.4\pm0.7$&
$14.6\pm1.4$&
$0.6\pm0.7$&
$6.8\pm3.5$&
$2.8\pm0.3$&
93&\\
$B$   &2.64& 
$5.8\pm1.2$&
$9.5\pm0.7$&
$19.6\pm2.1$&
$34.7\pm6.8$&
$2.1 $&
$11.0\pm1.1$&
$0.9\pm0.3$&
$5.7\pm0.7$&
$1.9\pm0.3$&
92.6&\\
$uvw1$&2.88& 
$6.1 $&
$9.8 $&
$14.5 $&
$22.5 $&
$3.9$&$11.4\pm0.3$&$ 0.8$&$6.0$&$3.1\pm0.5$&
86&\\
joint\,$^2$ & 0.81 & 
$6.5\pm0.2$&
$9.2\pm0.1$& 
$19.3\pm0.7$&
$38.3\pm2.7$& 
$3.6\pm0.3$& 
$15.4\pm0.5$&
$1.2\pm0.3$&
$5.4\pm0.4$&
$2.2\pm0.2$&
320&\\
\hline
\end{tabular}
\begin{tabular}{@{}llcr}
(1) & We fit {\tt mag = constant - s$_i$ * log$_{10}$(time)}, beginning at the start time with slope s$_a$ up to break time t$_a$,\\
& then continuing with s$_b$, until t$_b$, etc.. \\
(2)&The joint fit of I (OGLE) I$_c$ R$_c$ V, B (SMARTS) $uvw1\ \ uvm2\ \  uvw2$ (UVOT); $uvm2$ extends to 320d\\
(3)& normalised at t=1.0\,d the constant for each band is I=11.26, I$_c$=11.36, R$_c$=11.37, V=11.67, B=11.70, \\
&$uvw1$=11.43; all with $\pm 0.15$~mag, and $uvm2=10.95\pm0.35$, $uvw2=11.66\pm0.27$ mag.\\
\end{tabular}
\end{table*}

\subsection{\textit{Swift} UVOT photometry}

\Sw UVOT obtained UV photometry from day 2.88 until day 14, from day 24 to day 86, and from day 315 to day 320. 
The photometry was processed using the UVOT {\sc ftool uvotproduct} and the 20160321 version of the \Sw {\sc CALDB}. 
In addition a verification was made to eliminate observations that fell 
on areas of reduced sensitivity using the provisional small scale sensitivity 
map for UVOT\footnote{\url{http://heasarc.gsfc.nasa.gov/docs/heasarc/
caldb/swift/docs/uvot/uvotcaldb\_sss\_01b.pdf}}.
The \Sw UVOT filters have central wavelengths (on the AB system) of $uvw2$ =  1991\AA, $uvm2$ = 2221\AA, and $uvw1$ = 2753\AA~. The filter curves and a comparison of UVOT zeropoints in  AB and Vega photometric systems can be found in \citep{breeveld2011} and the \Sw CALDB.

The UVOT photometric observations in the $uvw1$, $uvm2$, and $uvw2$ 
filters are shown in Figure~\ref{fig_uv_opt_ir_lc}.
The light curve is discussed further in Section~\ref{sec:lc}. 
After day 38 there could be a possible significant contribution to the UVOT aperture from the two nearby sources. 
To reduce that contamination to the photometry we have used a $2\farcs5$ aperture, rather than the standard $5\arcsec$, with an aperture correction for count rates less than 0.5\,count\,s$^{-1}$ ($uvw1 < 19.72$~AB\,mag). 
The nearby stars, which fall within the UVOT PSF, are faint in the UV and 
no evidence of flattening of the UV light curve is found.

\subsection{The slope and breaks in the light curves }
\label{sec:lc}
The light curves are shown on a logarithmic time scale to illustrate the temporal breaks in Fig. \ref{fig_uv_opt_ir_lc} while  the rapid decline is more apparent in the linear plot, see Fig. \ref{fig_ogle_lc}. 
The light curve appears to consist of a few sections each with its own power law. 

For the $I$, $R$, $V$, and $uvw1$ light curves we fit a function linear in magnitude, but logarithmic in time, which is equivalent to a power law fit of flux(time) since magnitude is essentially a logarithm of the flux: $m = c - s\times log_{10}(t),$ with $c$ a constant, $m$ the magnitude, $s$ the slope, and $t$ the time. 
We do that initially on each section and band, find the break times from the intersections of the power laws, whereafter we repeat the fit per section. 
The intrinsic variability, see Fig.~\ref{fig_uv_opt_ir_lc}, which is also present during quiescence, makes fitting any smooth curve difficult; the measurement errors are much smaller than the variability. 
There appears to be a slight change in the light curves going from the red $I$ band to the ultraviolet, particularly noticable in the second slope $s_2$, but generally the variabilities are so large that assuming the same evolution takes place in all bands seems acceptable. The residuals of the fit for the $I$, $R$, and $uvw1$ bands  in Fig. \ref{fig_uv_opt_ir_lc}, middle panel, show visually the slight differences as compared to the overall evolution.
The overall light curve shows several well-separated intervals, and the joint fit in Table~\ref{tab:table_lc_slope} is a good representation for those. 
A possible explanation of the similar light curves from the UV to the IR is that the ejecta and possibly a central source are evolving as a whole because the ejecta are optically thin, while the source remains at a nearly constant temperature. 
This is surprisingly different from the colour evolution in CN V959~Mon \citep{page2013} which forms dust and in RN V745 Sco \citep{page2015} which has a red giant secondary. The main difference may be the size of ejected mass.

\citet{2010ApJS..187..275S} has pointed out that a large number of (possibly all) RNe show light curves with distinct plateau phases like we see in N LMC 1968, and suggests that as a defining characteristic of RNe.  
\citet{2014ApJ...788..164P} used this in a later paper in an attempt to uncover missed RNe. 
The plateau onset often coincides with the unveiling of the SSS. 

We combined all available 2016 and 2010 $V$ light curves from day 2--6 to estimate the rate of decay.  When extrapolating the fit back to the estimated eruption times we obtained values of $V$=12.3$\pm$0.5 at eruption, and derive $t_2 = 4.6\pm 0.5$d and $t_3 = 7.0\pm1.0$d, consistent with \citet{munari} and slightly slower than in $I$.

\begin{figure}
\includegraphics[width=\columnwidth]{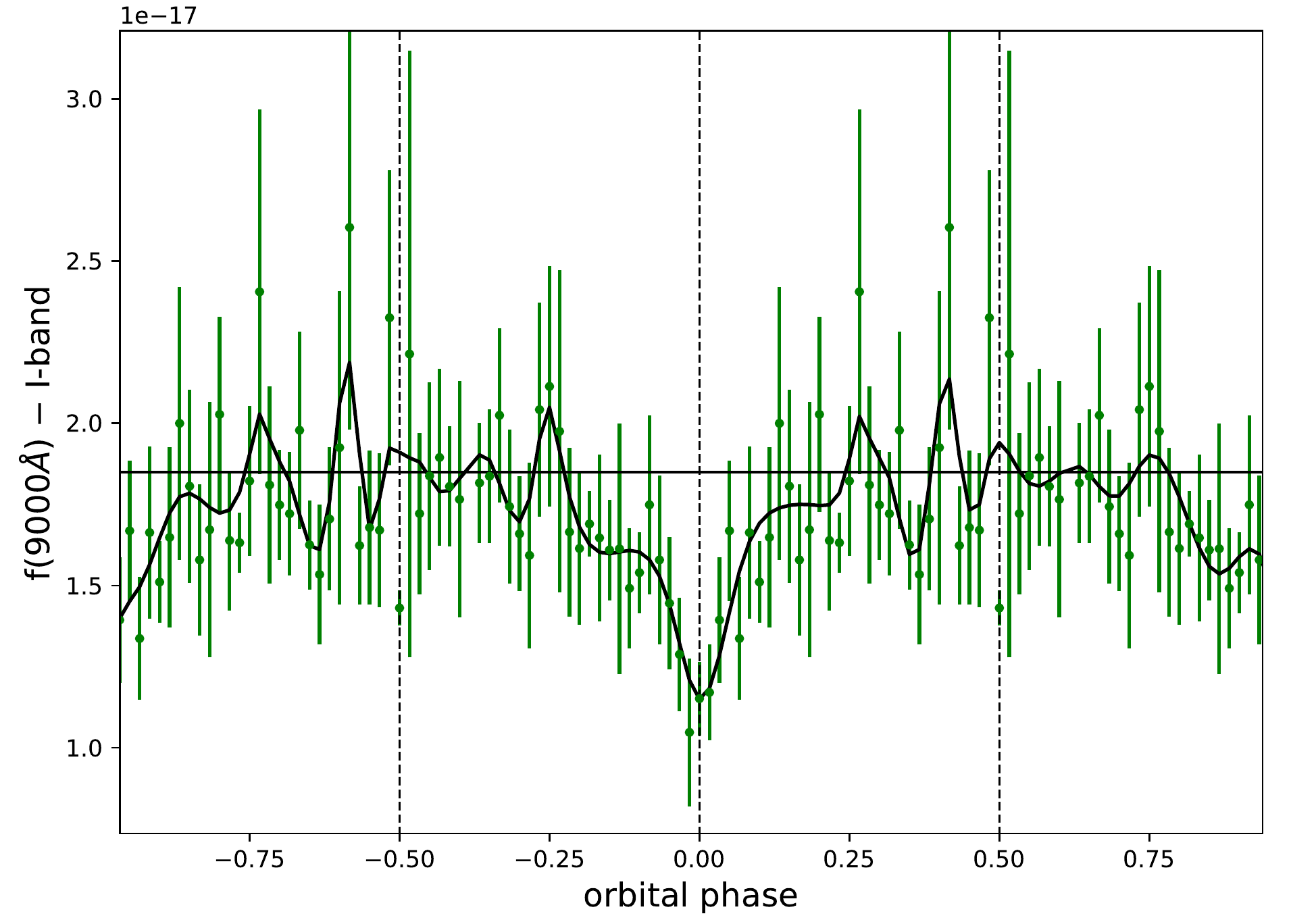}
\caption{
The I-band data folded and binned over the orbital period. The horizontal line is the mean flux for phase 0.25--0.75. The B-spline fit suggest an initial drop in brightness at phase -0.2, a further drop at phase -0.07, and a depth of the occultation of 0.40. 
}
\label{fig_IVBw1_phase}
\end{figure}

\begin{figure}
\includegraphics[width=94mm]{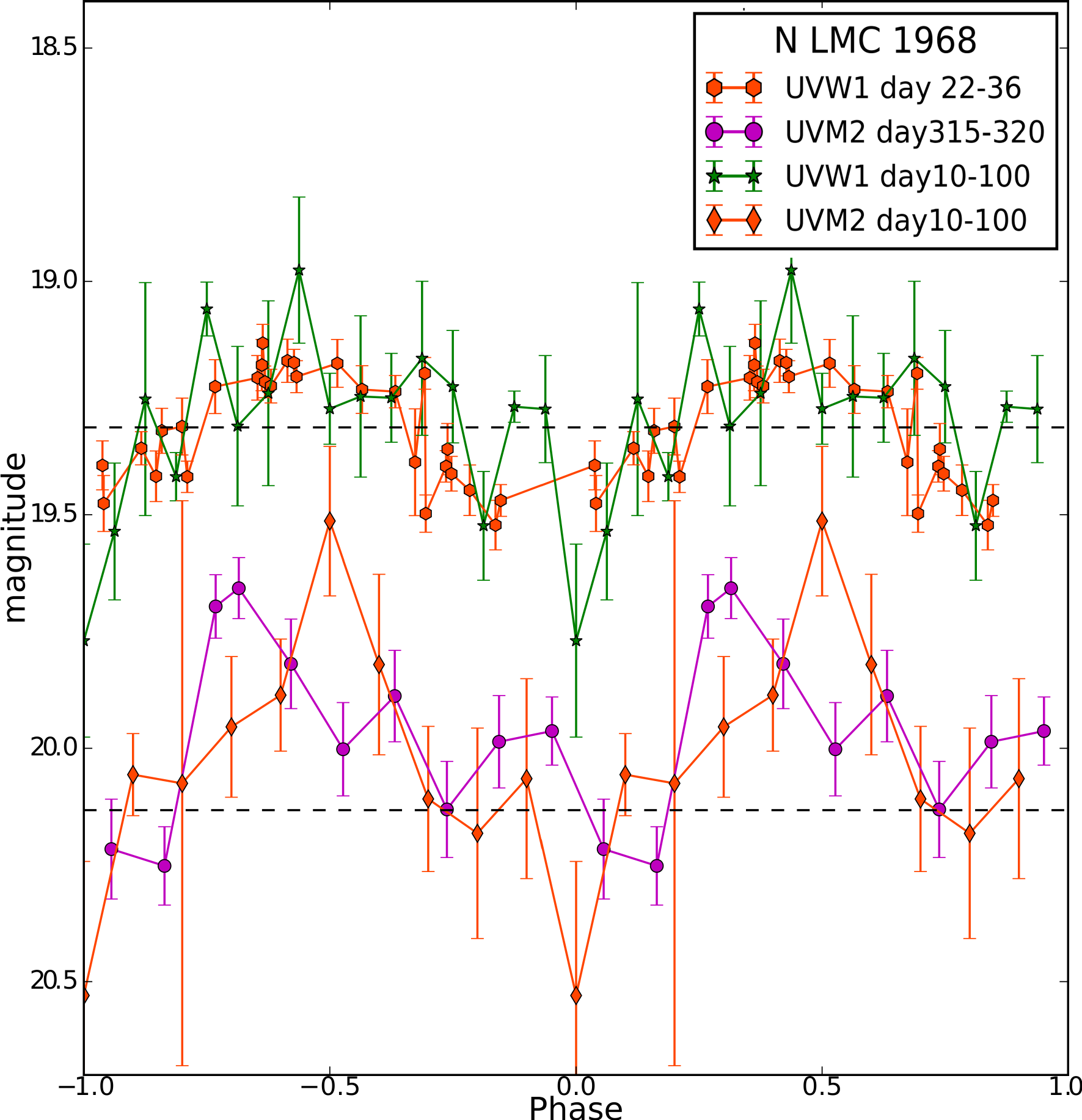}
\caption{Light curves folded on the period of 1.264d. On the top the average $uvw1$ from days 10--100 has been compared with day 22--36 (the time the soft X-ray emission peak). The minimum is shallow and broad. 
The bottom two curves compare the average for the $uvm2$ (2246\AA) band from days 10--100 compared with the light curve from day 315--320 after the eruption. The differences are due to variability.
The dashed line is the average magnitude for each period.}
\label{fig_m2_day22-36_315-320_phase}
\end{figure}

\begin{figure*}
\includegraphics[width=15.2 cm]{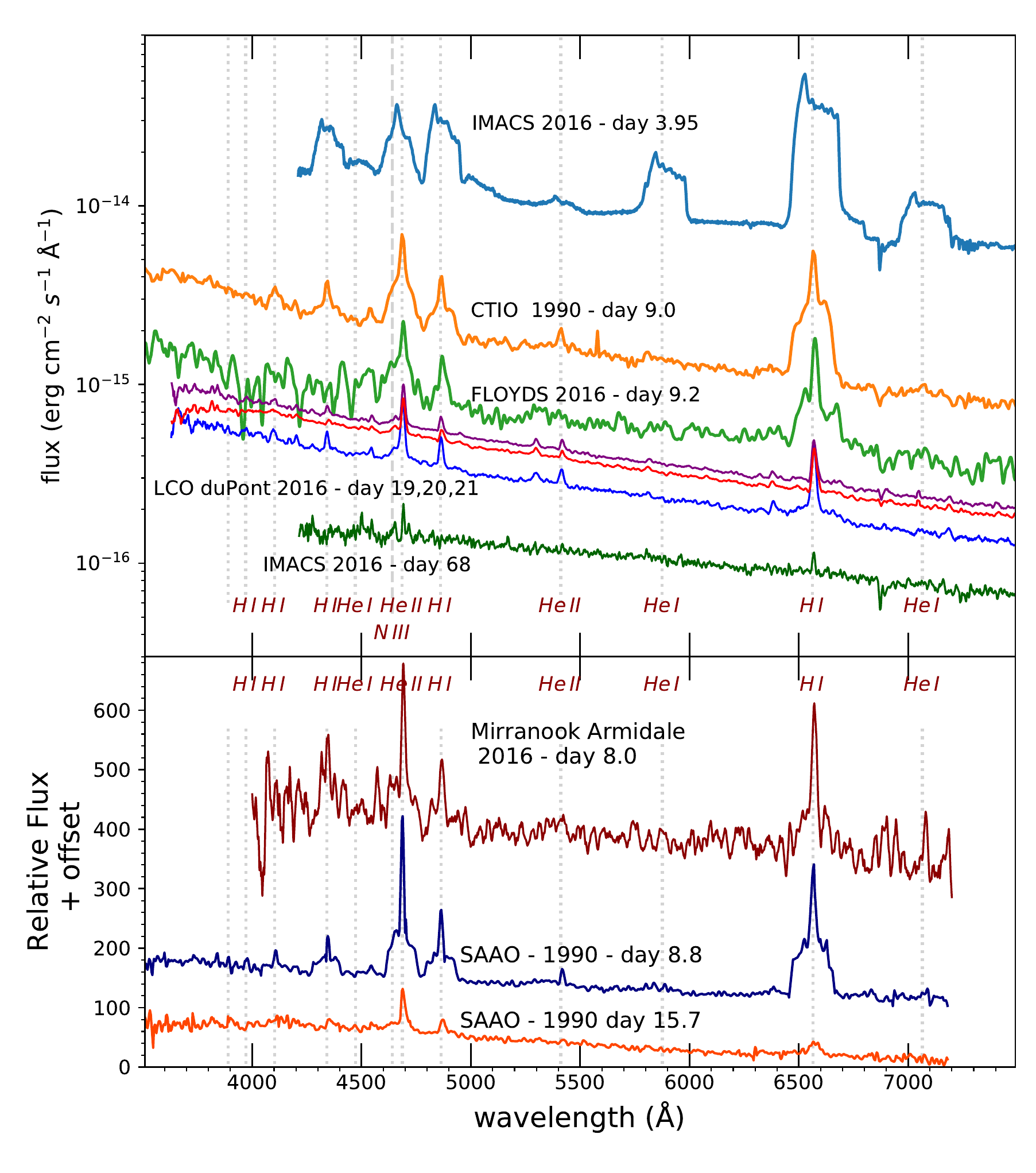}
\caption{The optical spectral evolution of the nova in 1990 and 2016. 
The spectra in the top panel have a flux calibration, those in the bottom panel are not flux calibrated. We show the optical spectra from the 1990 and 2016 eruption.
}
\label{fig:optical_spectra}
\end{figure*}

\subsection{Periodic photometric variability}\label{orb}
\label{section_periodic_photometric_variability}

\label{sec:orbit}

The OGLE project also monitored the nova (2010--2016) whilst in quiescence. 
Its quiescent mean magnitudes are $<I> = 19.29$\,mag, $<V> = 19.70$\,mag and colour $<V - I> = 0.41 \pm 0.06$\,mag. 
Furthermore, a periodic variation \citep{sekiguchi} is seen which is interpreted as being caused by the orbital variations \citep[see][for a periodogram]{mroz}. 
Henceforth we will adopt that interpretation. 
We obtained an  optimal period of 1\fd264329 using the analysis of variance (AOV) statistic method  \citep{schwarzenberg-czerny} using the 2010--2016 OGLE photometry, excluding the eruptions.   
The ephemeris was subsequently calculated using the O--C (observed $-$ calculated) 
diagram. The cadence of the observations, at 1--2 days, is close to the orbital period, limiting the accuracy of the ephemeris.

The solution for the ephemeris for the primary minimum in 
the $I$-band is 
\begin{eqnarray} 
\label{eq1}
\mathrm{HJD}_\mathrm{ecl} =  2455058.323 \pm 0.090 \\
+ \left(1.264329 \pm 0.000019\right) \times N  \   \nonumber  
\end{eqnarray}
where the errors are 1-$\sigma$. The main eclipse duration is about $\pm0.07$ in phase and about 0.6 mag deep (a 40\% decrease in flux) in the $I$-band, see Figs.~\ref{fig_IVBw1_phase} and \ref{fig_m2_day22-36_315-320_phase}.  The short deep eclipse would be consistent with the WD being eclipsed. 

We used the AOV method with the available data from OGLE and SMARTS starting from 23 days past the 2016 eruption (see Table~\ref{table_eruptions})  to determine if a period change could be detected, but we were unable to detect significant 
changes down at the $\pm0\fd003$ error level. 

Given the orbital period as derived from the $I$ band and using the ephemeris  from Eq.~\ref{eq1} we binned and folded our other photometry. 
We combined the $V$- and $I$-band photometry from OGLE and SMARTS, removing the first 50 days after the eruptions, leaving 559 data points. 
For $V$ (107 data points) and $B$ (64 data points) we also included SMARTS data from day\,10  after the rapid decline stopped, while removing the trend. For the $R$ (52 data points), the UVOT $uvw1$ (58 data points), $uvm2$ (25 data points) and $uvw2$ (14 points) bands we used data from the period of 10 days until 100 days after the 2016 eruption, again removing the trend. However, the average depth and width of the minimum are only well determined in the I band, and appear of similar depth in the other bands. 

In December 2016 (day 315--320) we obtained, during 4 \Sw observations of the nova system, ten exposures in the UVOT $uvm2$ band only. 
The phased $uvm2$ light curve shows an asymmetry but that is due to variability during the observation. 
This has been illustrated in Fig.~\ref{fig_m2_day22-36_315-320_phase} where the data are for day 46--89 (labeled day 10--100). 
The variability in  $uvm2$ appears to be mostly gradual and progressive, suggesting there slow changes in the occultation of the inner system continue long after the WD luminosity returned to the inter-eruption value. 

In the $uvw1$ the phased light curve for day 22--36 the occultation starts around phase -0.2, similar to the initial drop seen in the I-band B-spline fit.   
While the SSS phase was still on-going the $uvw1$ shows nearly the same sinusoidal profile as the long term average which suggests the presence of an accretion disk by day 28, see Section \ref{sec:accretion}.



\subsection{Optical Spectroscopy}
\label{sec:optical_spectroscopy}
We obtained a series of spectra at Las Campanas Observatory (LCO) of the 2016 eruption of N\ LMC\,1968 starting with a 120s exposure on 2016 Jan 24.0345 (JD\,2457411.5345, day 3.95) 
using the IMACS Short-Camera instrument on the 6.5m Magellan-Baade Telescope \citep{dimille}. We obtained  600s+2$\times$1200s long exposures on 2016 Feb 8, and 3x1200s on 2016 Feb. 9 and 10 (days 19--21) on the 2.5m du Pont telescope using {WFCCD/WF4K-1}, and finally exposing for 900s on Mar. 29 (day 68) with the IMACS Short-Camera, see Table~\ref{tab:all_spectra}. Spectral resolution is $\sim5$\AA~ for the IMACS data, and $\sim8$\AA~ for the WFCCD.
We used standard IRAF routines to reduce the LCO spectra of the nova as well as those of spectrophotometric standard stars observed on the same nights for the flux calibration.  

The first 2016 spectrum shows a moderately blue continuum dominated by broad Balmer,  \ion{He}{i} (triplet) and \ion{He}{ii} emission lines, see Figure~\ref{fig:optical_spectra}. 
The lines have a FWZI of about 10,000\,km\,s$^{-1}$, implying velocities of 5000~km~s$^{-1}$ and present jagged profiles. 
A bright narrow emission peak at a velocity of $\sim1600$\,km\,s$^{-1}$ is clearly visible on the blue edge of the Balmer and 
\ion{He}{i} lines, but there is no narrow centered component on the lines as seen day 8 and later. 
The \ion{He}{ii} 4686\,\AA\ line appears to have a more symmetric 
profile, which is due to a blend with the Bowen \ion{C}{III} and/or \ion{N}{III} lines.

A spectrum (4000--7200\,\AA) was obtained  on 2016 Jan. 28 starting at 10:17 UT (day 8.0, phase 0.73) with 11x300\,s exposure at the Mirranook Armidale site using a LISA spectrograph on a C11 telescope with a 23 {$\mu$}m wide slit (about 3\arcsec~ on the sky).
The spectrum shows the H$\alpha$, H$\beta$ and He\,II~4687\,\AA~ lines, and the lines feature narrow components.  

Further spectra were obtained using the FLOYDS instrument\footnote{https://lco.global/observatory/instruments/floyds/} 
on the 2.0\,m Faulkes Telescope South at Siding Spring Observatory, NSW, 
Australia on 2016 Jan.\ 29.6 UT. (JD\,2457417.1, day 9.2, orbital phase 0.63).  
We obtained a series of spectra $3\times900$ seconds long. 
The spectral range covers 3200\,\AA\  to $1\,\mu$m at a resolution 
of $R\approx550$. 
The signal-to-noise of the combined spectrum is particularly low and only four emission 
lines are clearly visible, H$\alpha$, H$\beta$, \ion{He}{ii} (4687\,\AA), while the \ion{He}{i} lines are no longer seen.  
The H$\alpha$ line consists of a bright narrow central peak with 
FWHM $= 1900 \pm 100$\,km\,s$^{-1}$ on top of a broader pedestal with  
FWZI $\approx10,000$\,km\,s$^{-1}$. 
We calibrated the spectrum using the mean fit to the photometry given in Table \ref{tab:table_lc_slope}. 

In Fig.\,\ref{fig:optical_spectra}  the SAAO spectra from the 1990 eruption also show a centered narrow line component on day 8.78 and 15.73 similar to that seen in 2016, but the underlying broad "pedestal" of the line profiles shows a different profile in 2016 than in 1990.

Table \ref{tab:all_spectra} gives an overview of all the spectra, including the orbital phase.


\begin{figure}
\includegraphics[width=88mm]{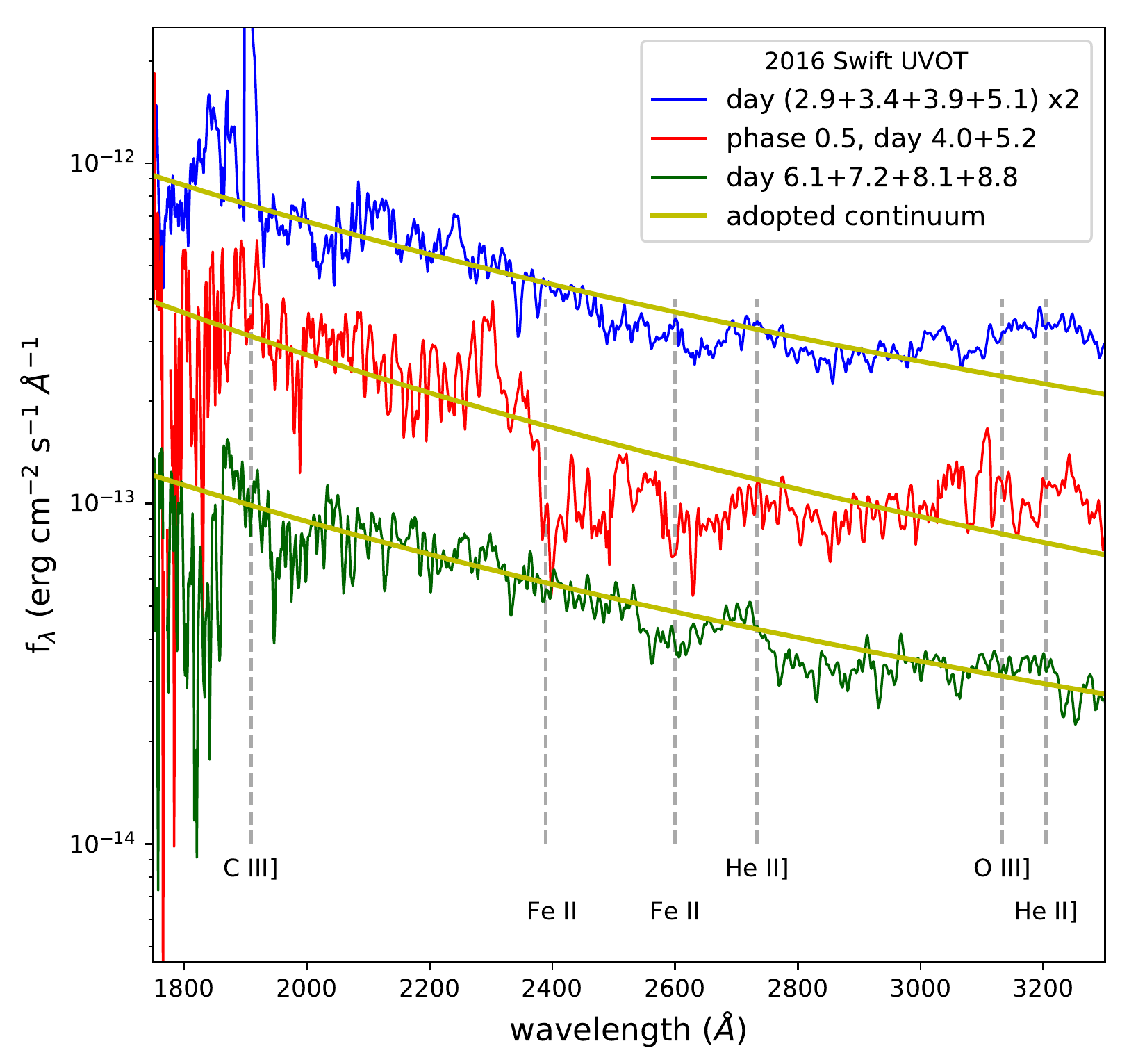}
\caption{
The observed Swift UVOT UV spectrum.  
The spectra were summed to improve S/N, weighted by the flux error. Since the early spectra are brighter, they will temd to dominate.  The spectra past day 6 were taken at an offset and do not suffer from second order overlap which raises the continuum level in the earlier spectra above 2800\,\AA. 
A power law continuum has been fit to illustrate the presence of Fe~II absorption features which are more prominent at phase 0.5, days 4--5.2. There are many Fe II lines in the 2380--3020\,\AA~  band;  the UV 1 and 2 multiplet locations have only been indicated.    
A reddening correction for E(B-V) = 0.07 was applied to the spectra , see Section~\ref{sec:reddening} and the earliest spectrum was multiplied by two for clarity. 
The spectra show lines of He\,II 2734, 3204, O\,III 3133, and C\,III] 1909\AA.
}
\label{fig:uvot_spectra}
\end{figure}

\begin{figure*}
\includegraphics[angle=90.,width=152.0 mm]{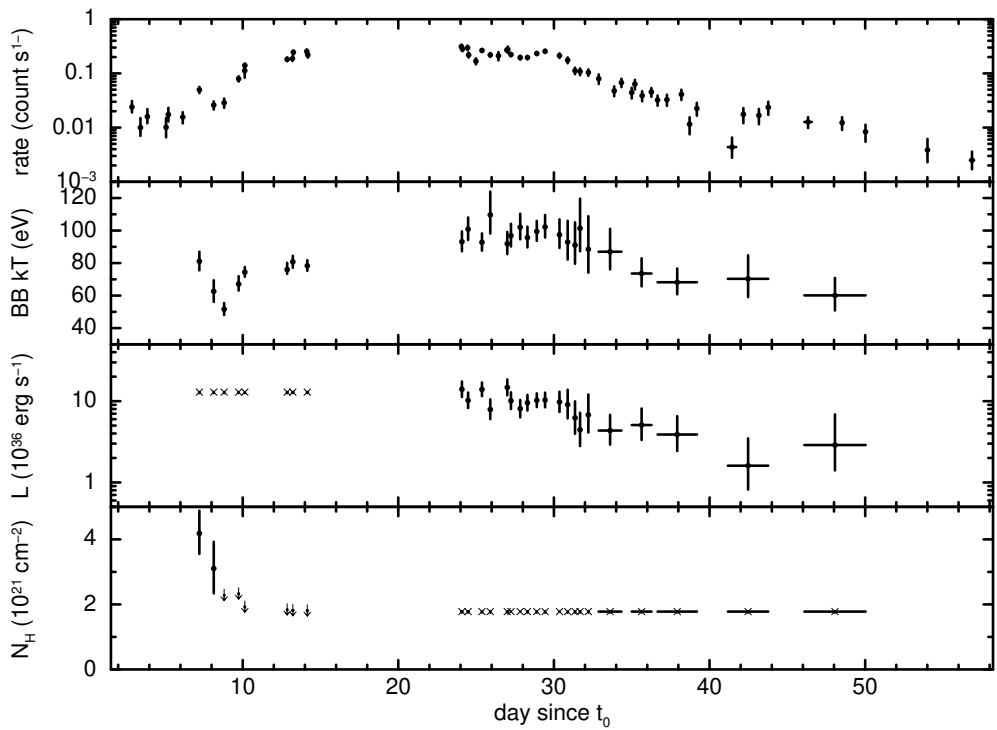}

\caption{Top panel : The observed light curve over the full XRT band (0.3--10\,keV). 
Second panel: The temperature of the blackbody fit to the soft spectrum.
Third panel: For times > 14 d: The estimated bolometric luminosity of the BB, assuming a distance of 50 kpc; for early times L$_x$ has been set to the mean from day 20--30. 
Fourth panel: Prior to day 14: the estimated NH column. After day 14: the adopted NH$_{ISM}$ value based on a fit to the spectra past day 20.}
\label{fig:xrt_lc_BBfit_Lbol}
\end{figure*}

\subsection{\textit{Swift} UV spectroscopy}


Daily \Sw UVOT UV grism spectra were obtained between 2016 Jan.\ 23~06:40 UT (day 2.875) and 2016 Jan.\ 29  (day 9).

The first grism observations consisted of two 500\,s segments with different roll angles in order to have a different zeroth order contamination from field stars. 
The details of the UVOT spectra have been given in Table \ref{tab:all_spectra} and \ref{table_grism_obs}. 
The spectra were processed using the calibration of \citet{kuin2015} and code described in \citet{kuin2014}.
The calibration used includes the  2017 update to the sensitivity loss which 
affects the spectra below 2000\AA.

After extraction, the UVOT spectra were validated since the UVOT grism images need careful analysis  \citep[see ][ for details]{2018A&A...619A.104S}. 
The grism images were compared to the star field and contaminating zeroth orders were flagged. 
This was especially important for the first six spectra because no offset on the detector was used (see Table~\ref{table_grism_obs}). \footnote{The advantage of using an offset position on the detector around (1000,1600) is no second order overlap while zeroth order contamination is restricted to below 2000\,\AA. }
Though the UV grism spectrum from day 6.14 (phase 0.22) was unusually bright when compared with the longer term trend, it was found to be consistent with the photometry taken right before and after the spectrum showing  that a brightening was taking place at that time.  
In order to correct for errors in the wavelength fiducial point \cite[the anchor point, see ][]{kuin2015} which applies equally to all spectra on a grism image, spectra of a bright nearby F5 star were extracted and used to determine a correction.

The spectra were very noisy, so we summed them to enhance the UV line emission. However, it was clear that spectra near phase 0.5 were different, while the other spectra were essentially the same, so these were summed separately.
In Fig.\ref{fig:uvot_spectra} the spectra near phase 0.5 have been shown together with earlier and later summed spectra. 
We found weak broad lines of of He\,II 2734 and 3204\,\AA, consistent with the optical spectra. 
A bright line appears to be present at $\sim$1909\,\AA~  due to C\,III].
We were unsure if the loss of flux below 1850\AA~ in all spectra is due to absorption or to the calibration of the sensitivity loss which is particularly difficult where the response falls to zero. We also noted a broad line which we identify as the Bowen O\,III 3133\AA~ line which is pumped by He\,II, consistent with the appearance of the C~III/N~III Bowen lines at 4630--4650\,\AA. 
Since we do not see strong UV line emission, the ejecta are not in the nebular phase before day 9.

\begin{figure*}
\includegraphics[angle=270,width=152 mm]{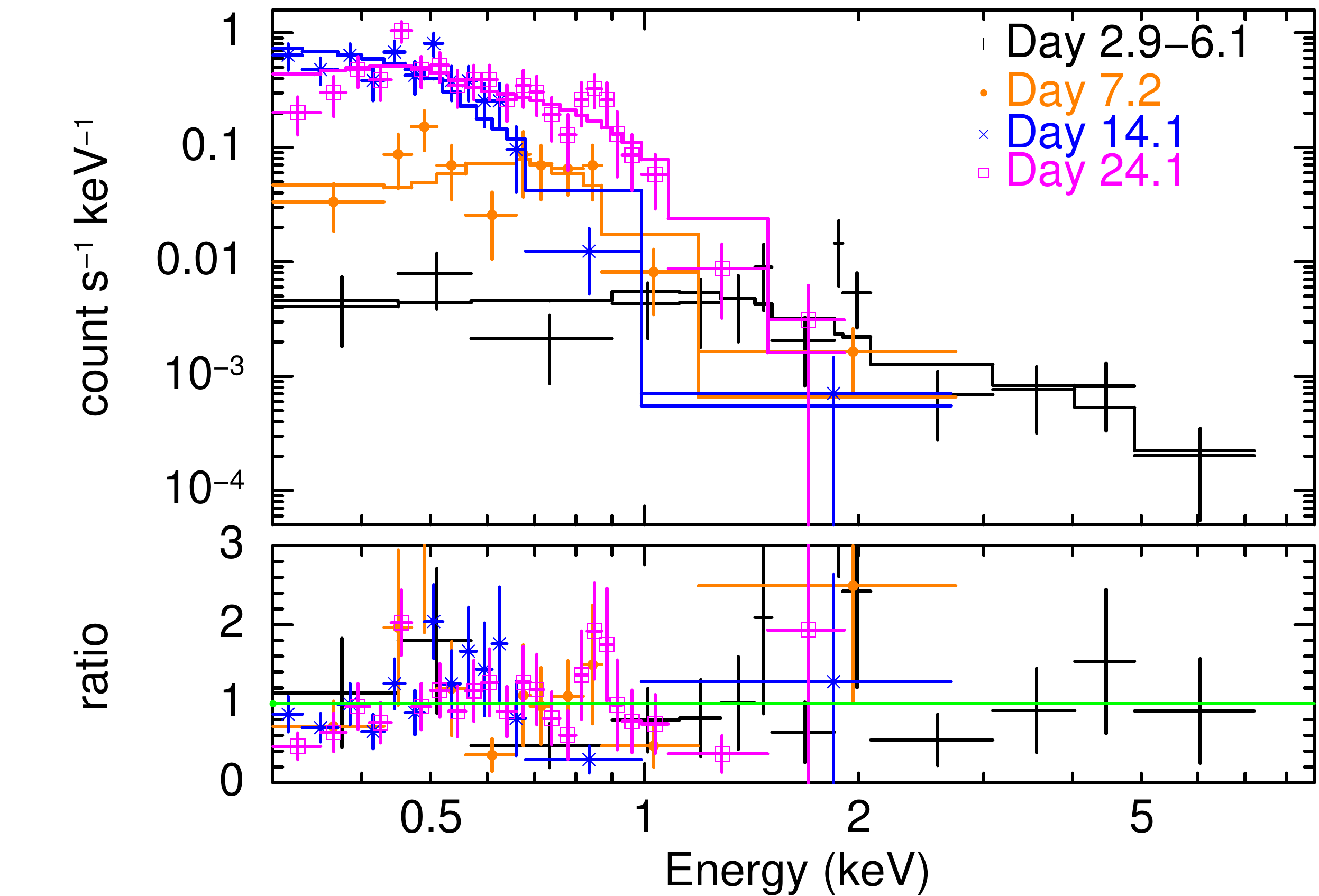}
\caption{A sample of the Swift-XRT spectra fitted with a BB model. 
The lower panel shows the ratio between the data and model.}
\label{fig_xrt_spectra_and_BBfit}
\end{figure*}

\subsection{\textit{Swift} XRT spectra} 

\label{section_xrt_obs}

Spectra were extracted for each individual snapshot of XRT data (where a snapshot is a continuous \Sw pointing) after the soft emission became evident (with the first good soft X-ray spectrum for day 7.2), except for data after day 33, where several observations were combined to get spectra of sufficient signal-to-noise. 
A single spectrum was also extracted for the early, pre-super-soft-source (SSS) emission (days 2.9--6.1). The spectra were binned to a minimum of 1\,count\,bin$^{-1}$ in order to facilitate fitting using the \citet{cash1979} statistic within {\sc xspec}.

\label{sec:xrtspectra}

We can expect that prior to the rise in SSS emission on day 6 the ejecta are optically thick for X-ray emission from the WD below, and the hard X-ray spectrum is due to the optically thin emission in a shock in the ejecta.  After day 6 the ejecta become transparent and the WD photospheric emission shows through. Between day 6 and the time that the X-ray emission peaks around day 14 we expect the column density to include a decreasing contribution from the ejecta above that from the ISM.  At late times, absorption from the ejecta will be negligible, and we can use that to derive the ISM column N$_{\mathrm{H}-ISM}$ from modelling our XRT spectra. 

We start with determining the ISM \NH~ by modelling the late time spectra, day 21 and onwards. 
From our late time models we derive  $N_\mathrm{H}$ = $1.8\times10^{21}$\,cm$^{-2}$. 
We can compare this to the LAB \citep{kalberla} 21\,cm survey, separating the LMC and Galactic components, and using the total velocity range, $-150$\,km\,s$^{-1}$ to  $+300$\,km\,s$^{-1}$ the column $N_{\mathrm{H}{-ISM}}$ = $1.8\times10^{21}\,\mathrm{cm}^{-2}$ within a $0\fdg27$ beam\footnote{\url{https://www.astro.uni-bonn.de/hisurvey/profile/index.php}}. 
The model value is thus consistent with the interstellar one.

The early, day 2.9--6.1, spectrum was fitted with an optically-thin thermal component (APEC), with $kT = 5.5^{+19.1}_{-2.6}$ keV, and $N_\mathrm{H}$ fixed at the late-time value. 
The 0.3--10\,keV unabsorbed luminosity of this early hard component is $1.9^{+0.7}_{-0.5}\times10^{35}$\,erg\,s$^{-1}$ which is lower than the (later) soft X-ray luminosity. 
The model has been chosen as appropriate for  shock-heated plasma emission 
from internal shocks and was assumed to be unabsorbed by the ejecta.

During day 6--14 we observe a rise in the XRT count rate due to the increase in the soft component. 
Since there are very few counts above $\sim1.5$\,keV after day 6, a model consisting of an absorbed blackbody (BB) component was sufficient to parametrise the SSS spectra. 
While a WD model atmosphere, such as the T{\" u}bingen Non-Local Thermal Equilibrium Model Atmosphere Package (TMAP\footnote{\url{http://astro.uni-tuebingen.de/~rauch/TMAF/flux_HHeCNONeMgSiS_gen.html}}) would be more physically appropriate, the temperatures required by these XRT spectra were typically too high for these model grids.
We assume that the SSS emission originates from the hot WD photosphere and that the rise in SSS is due to the ejecta becoming transparent as proposed by \cite{1996ApJ...456..788K} and \cite{shore1996}.
We make the further assumption that the WD X-ray luminosity is constant from the eruption until the end of the SSS phase \citep{1996ApJ...456..788K}. 
We find the bolometric luminosity of the soft X-ray source by using  the mean value during the SSS plateau from a fit with \NH~ fixed to the ISM value during day 20--30, i.e., L$_{X-Bol} \approx$ 1.1$ \times 10^{37}$ erg s$^{-1}$, and use that for the model prior to day 14 when \NH~ is 
high\footnote{The modelling choices made here are made because the fitted parameters are not completely independent, and optimising just on the goodness of fit may result in unphysical values}. 
During day 6--14 we assume \NH~ includes an additional component due to the optical thickness of the ejecta and fit a blackbody model leaving \NH~ to vary with a lower limit set by the interstellar medium value \NH$_{-ISM}$. 
The result is shown in Fig.~\ref{fig:xrt_lc_BBfit_Lbol} where we see that indeed the column density shows a steep decrease from \NH~{$> 4\times 10^{21}$ cm$^{-2}$} to the ISM value over approximately a three day period prior to the SSS ramp up.  
Using the fitted blackbody temperature ($\sim$~100~eV) and the fixed luminosity (1.1$ \times 10^{37}$ erg s$^{-1}$), we derive a BB radius of $\approx$3270~km for the hot photosphere of the WD during the peak SSS emission. There are some caveats: (1) the assumption of a BB spectrum may not be valid and the temperature may not have the usual physical meaning; (2) the system is eclipsing, so the rim of the accretion disk may hide part of the WD emission; and (3) the theoretical radius of a 1.3~M$_\odot$ WD is smaller than 2800~km \citep{carvalho2018}, which suggests the error in the blackbody temperature estimate is of order 8\%.

Figure~\ref{fig_xrt_spectra_and_BBfit} shows a comparison of four 
spectra obtained throughout the rise to peak count rate. 
The first spectrum, up to day 6, is hard, as described above.
On day 7.2 the ejecta have started to clear though the value of \NH\, still exceeds the ISM value. 
Despite the count rate being very similar before and after the pole constraint gap, days 14.1 and 24.1 show very different spectral shapes. The BB fit shows the later spectrum being about 20\,eV hotter. 
That suggests that the radius of the WD photosphere decreased by about 30\% between day 14 and day 24.  

In Fig.~\ref{fig:xrt_lc_BBfit_Lbol} the fitted luminosity is found to be an order of magnitude lower
than the Eddington luminosity for a $1.3\,M_{\sun}$ WD,  which is $1.6 \times 10^{38}$\,erg\,s$^{-1}$. 
The spectral hardening on day 24.1 takes place at the end of the plateau in the optical light curves which break around day 19.3, see Table~\ref{tab:table_lc_slope}. 
Assuming a blackbody spectrum for the SSS source the soft X-rays do not contribute significantly to the UV-optical emission. Therefore, to derive the bolometric luminosity the contemporary UV-optical/IR emission needs to be added to the $L_x$ from the BB fit. We come back to the luminosity in Section~\ref{sec:uv_xray_ratio}.


%
\subsection{IUE, CTIO and SAAO data of the 1990 event}

\begin{figure*}
\includegraphics[width=162.mm]{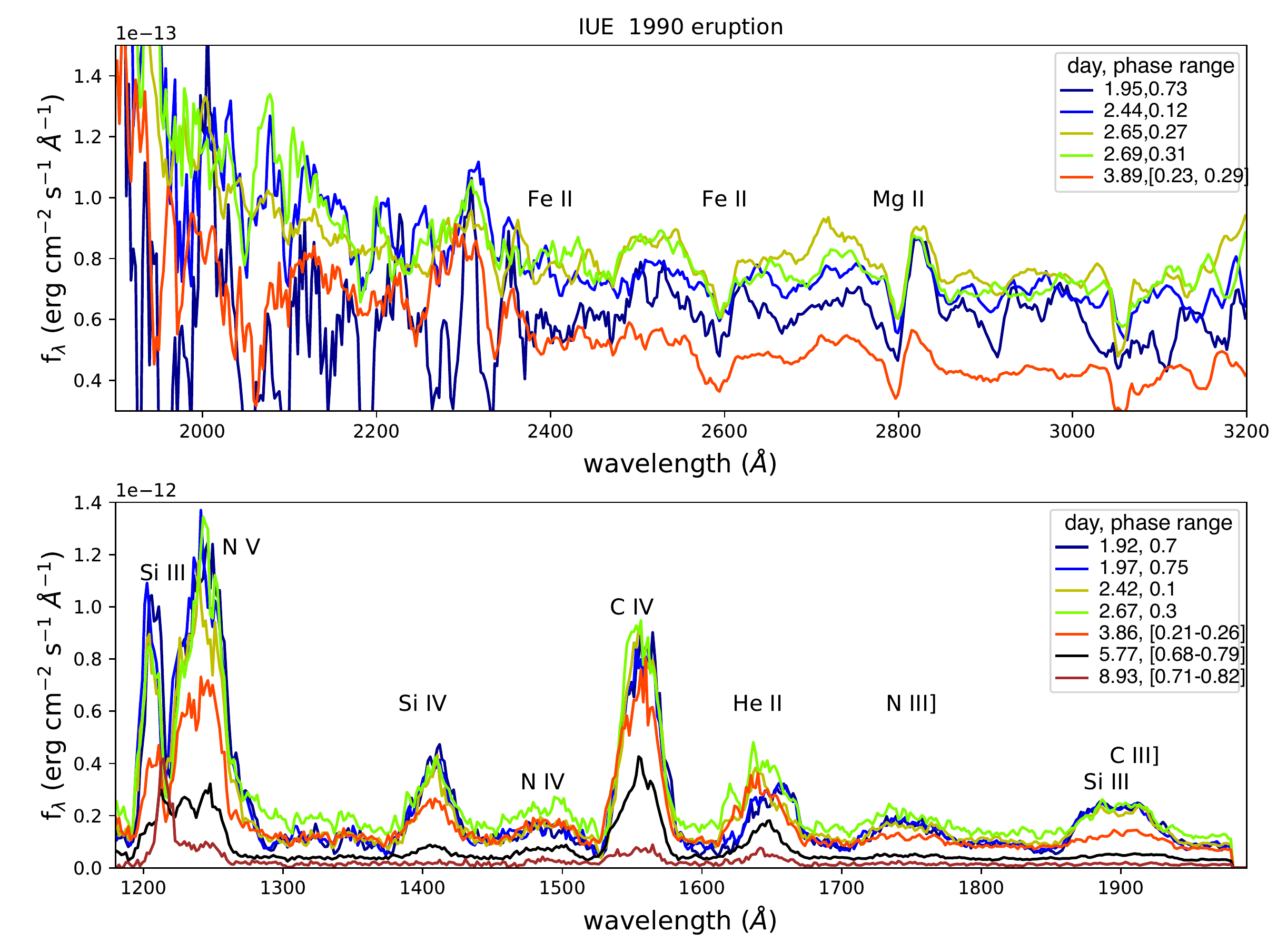}
\caption{The IUE long (LWP) and short (SWP)  wavelength spectra from the 1990 eruption. Phase is from our ephemerides. The LWP spectra show absorption of Fe~II and Mg~II.  The LWP emission lines have been indicated.  No reddening correction was applied. }
\label{fig:iue_swp}
\end{figure*}

The International Ultraviolet Explorer ({\sc IUE}) spectra we retrieved are from the IUE Final Archive which improved upon the earlier spectral extraction \citep{1996AJ....111..517N}, see Fig.~\ref{fig:iue_swp}. The long wavelength (LWP) IUE spectra cover 1900--3200\AA, comparable to the UVOT UV grism, but suffer from much noise and weak emission lines. The strongest line often seen is \ion{C}{iii]}. The shorter IUE wavelength band (SWP) contains strong emission lines from C, N, and O, and H~Ly$\alpha$.  
For the IUE spectra, as well as the spectrum of N\ LMC\,1990b taken with the 1.9\,m telescope at the South African Astronomical Observatory (SAAO) in Sutherland and for the 
spectrum which was taken at the Cerro-Tololo Interamerican Observatory (CTIO)
we can use the OGLE ephemeris from Eq.~\ref{eq1} for attaching an orbital phase to the 1990 outburst observations with an uncertainty of $\approx$0.15 in phase derived from the uncertainty in the ephemeris, see Table~\ref{tab:all_spectra}. 
With this uncertainty in the assigned phase for the 1990 observations in mind, we proceed to use them in our interpretation of the nova.

\section{Observational results}
\label{sec:obs_results}


\subsection{The companion star}
\label{sec:secondary}

\cite{mroz} suggested that the period can in principle be twice that given above. However, the  UV spectrum at phase 0.5, day 4 and 5.2 showed an increase in Fe~II absorption. Though this is due to an ionisation effect, (see Section~\ref{section_L2outflow}), it could also be associated with extra mass present outside the L2 point. 
The orbital period suggests that for a Roche-lobe filling secondary, the secondary must be evolved \citep{shore,mroz}. 

We checked  the pre-eruption WISE IR data for insight into the  contribution from the companion. 
However, a comparison with the higher resolution OGLE image, see Fig.~\ref{fig_ogle_finder}, shows that there are two sources within $3\arcsec$ of the nova, less than the $6\farcs1$ resolution in WISE Band W1 and the  $6\farcs4$ resolution of Band W2 so that no useful information can be obtained on the companion from WISE data. 

The 2016 Feb. 19, 20 and 21 LCO spectra cover the 3800--9000\,\AA~ band at phase 0.98, 0.76 and 0.51, respectively. 
This was at the end of the plateau in the light curve. To adjust for the brightness changes we scaled the spectra using the flux of the H$\alpha$ line which is formed in the ejecta. We see at phase 0.51 the whole WD-facing atmosphere and at phase 0.76 only half. After scaling the spectra at phase 0.76 and 0.51 are identical within the error, so we do not observe any difference in emission due to the  heated atmosphere on the side of the secondary facing the hot WD. From these observations we derive a 3-$\sigma$ flux limit for the heated atmosphere in the secondary is less than $1.1 \times 10^{-13} {\rm erg~ cm}^{-2} {\rm s}^{-1}$ which is negligible. 

The slope of a power-law fit to the LCO spectrum on day 20 is $2.27 \pm 0.03$, close to what an $\alpha$-disk model predicts, supporting that the accretion disk was present.

\subsection{The mass of the WD}
\label{sec:mass}

The luminosity-temperature relations in both \citet{2005A&A...439.1061S} and \citet{2013ApJ...777..136W} show that a peak temperature of $>$100~eV implies a high WD mass of $>$1.3\,M$_{\odot}$. 
Compared to CN eruptions, RNe have much shorter intervals during which material can be accreted onto the WD, and less is needed to get ignition. Therefore, there is less material ejected during the eruption of a recurrent system. Super-soft X-rays will only be observable when the ejecta have become optically thin \citep[e.g.][]{1996ApJ...456..788K}, therefore RNe and, by extension, high mass WDs, are expected to have both short turn-on and turn-off times for their SSS phases. This is indeed found to be the case for N\ LMC 1968, with turn-on and -off times of about seven days and $\sim$30 days respectively. These times are completely consistent with the correlations found by \citet{2014A&A...563A...2H} when analysing a sample of M31 novae. The earliest detection of a SSS so far was for V745 Sco \citep{2015MNRAS.454.3108P}, with soft X-ray emission first seen about four days after eruption. V745~Sco has a recurrence time of $\sim$25~yr.
The rapid RN M31N 2008-12a also showed an early turn-on of the SSS phase, six days after the nova eruption \citep{2015A&A...580A..46H}. 
The small 3270km radius of the WD photosphere as derived from the blackbody model of the peak SSS XRT spectrum in Section \ref{sec:xrtspectra} is also indicative of a massive WD with M$>$1.25\,M$_{\odot}$ \citep{carvalho2018}.

\begin{figure}
\includegraphics[width=\columnwidth]{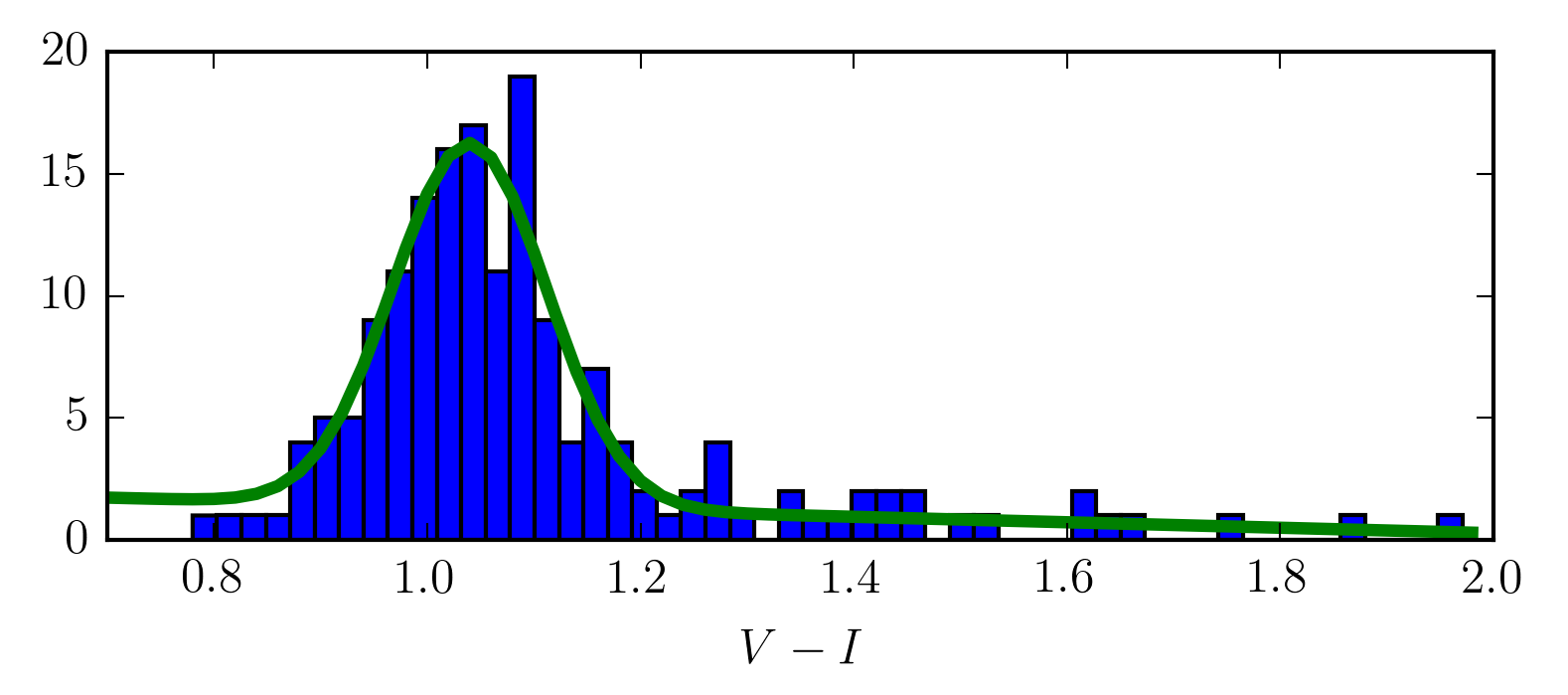}
\caption{The histogram of the $V-I$ colour for the red clump in the Colour-Magnitude Diagram of the $2\arcmin\times2\arcmin$ region around the 
nova.}
\label{red_clump}
\end{figure}


\subsection{Reddening}
\label{sec:reddening}

We considered several lines of evidence for determining the interstellar reddening toward the nova. 
We made an estimate of the reddening in the direction of the nova using the Red Clump \citep{redclump} in the Colour-Magnitude Diagram  by centroiding on a 2{\arcmin}{$\times$}2\arcmin\  region around the  nova ($30\times30$\,pc at a distance of 50\,kpc), where the measured colour $(V-I)_\mathrm{RC}$ = $1.02 \pm 0.01$ mag (see Figure~\ref{red_clump}). 
Assuming an intrinsic colour for the Red Clump of $(V-I)_\mathrm{RC}$ = 0.92 
for LMC metallicity, the reddening  is $E(V-I) = 0.10$\,mag. 
Assuming a Cardelli law \citep{cardelli} this corresponds to $A_V =  0.21$\,mag, $A_I = 0.10$\,mag and $E(B-V) = 0.07$\,mag. 
This is consistent with  NED\footnote{NASA/IPAC Extragalactic Database,  \url{http://ned.ipac.caltech.edu}, based on SDSS data} Galactic extinction calculator which gives a visual extinction of $A_V = 0.206$.
For an IR-independent estimate we take \NH~ discussed before in Section~\ref{section_xrt_obs} from the XRT spectral fit and the LAB 21\,cm survey; the column $N_\mathrm{H} = 1.8\times10^{21}\,\mathrm{cm}^{-2}$.
This value is consistent with the reddening and $N_\mathrm{H}$/E(B-V) ratio that has been reported for the LMC \citep[][]{Koorneef}.
Using the \citet{bohlin} calibration this is $E(B-V) = 0.09$.  
For the \citet{liszt} calibration for $|b|>20^\circ$ $E(B-V) = 0.07$. We adopt $E(B-V) = 0.07 \pm 0.01$ in this paper.

We can use the UV spectra to learn more of the reddening specific to this nova. Both the 1990 IUE LWP spectra and the \Sw UVOT spectra cover the $\lambda$\,2175\,\AA\  feature from which a lower limit to E(B-V) can be derived. Using the Verbunt method \citep{verbunt} after summing all the spectra and fitting the continuum, the \citet{cardelli} Galactic extinction law was applied for various values of E(B-V) using R$_V$ = 3.1 and visually inspected. The bump in the spectrum (positive or negative) disappeared for a very low E(B-V) = 0.006$\pm$0.003, suggesting the absence of the Galactic $\lambda\,2175$\,\AA~ feature in that direction.  


\begin{figure}
\includegraphics[width=\columnwidth]{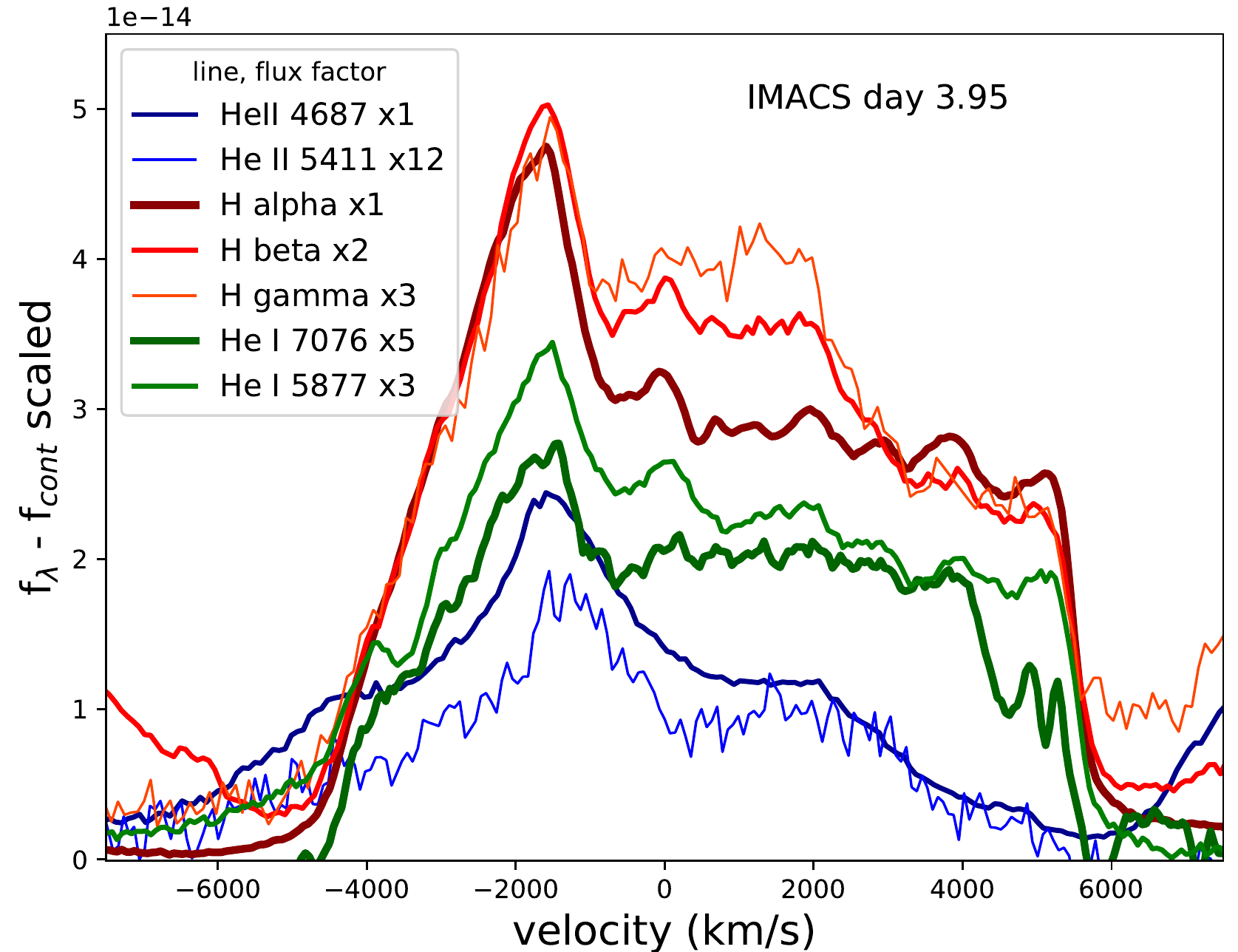}
\caption{On day 3.95  in the 2016 outburst, the H and He line profiles in the IMACS spectrum have been displayed in velocity space. The He II lines show less emission on the red wing than the H and He~I lines.  Assuming symmetry, the line H and He~I profile centers are at 1035~km~s$^{-1}$}
\label{fig:imacs2016_velocity}
\end{figure}


\begin{figure}
\includegraphics[width=\columnwidth]{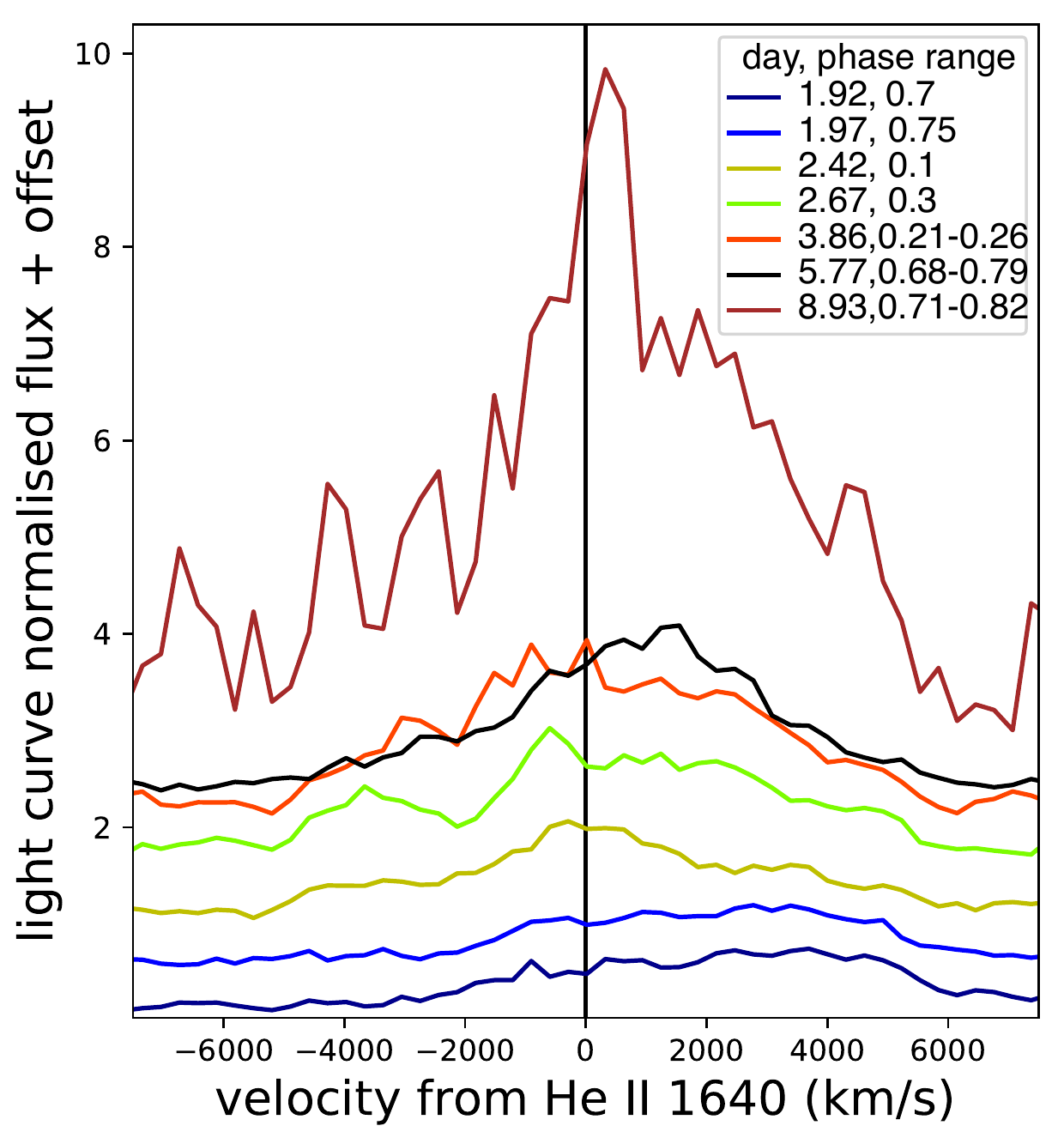}
\caption{The changing profile of the 1990 IUE SWP  He~II~1640 line profile up to day 9.  The flux has been normalised using the light curve fit in Table~\ref{tab:table_lc_slope} and offset for clarity. The He II flux did not drop as fast as the nova brightness, and thus the earliest normalised line is the weakest, and the latest is the strongest. Note that between day 5.77 and 8.93 a central narrow line component appears. Ignoring the narrow line profile, the broader line tends to the red for orbital phase near 0.7 and slightly to the blue for orbital phase near 0.2. These line profile changes are not seen in the spectra of neutral H and He.}
\label{fig:HeII1640}
\end{figure}


\begin{figure}
\includegraphics[width=\columnwidth]{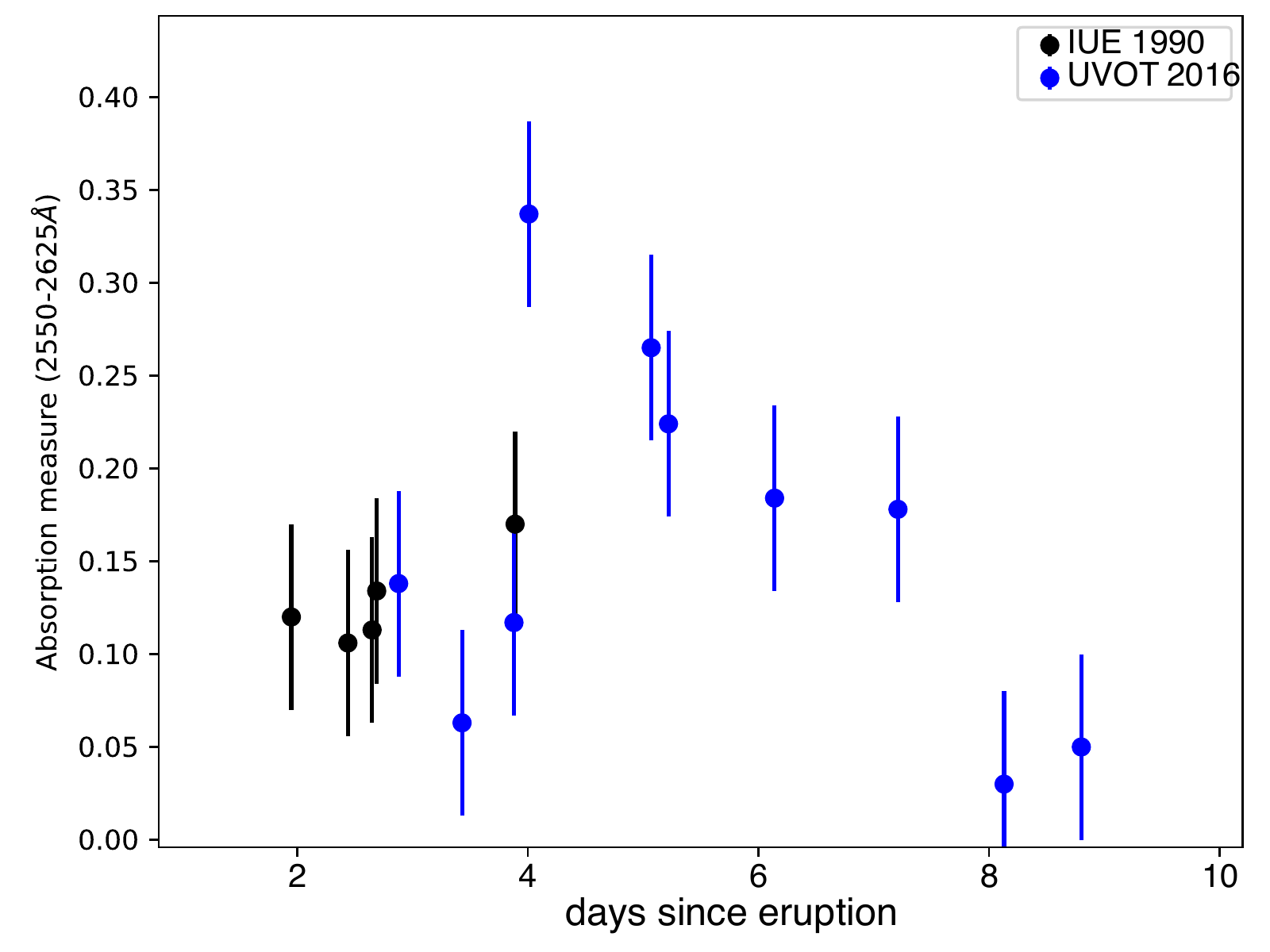}
\caption{The relative absorption below the continuum of the region 2550--2625\,\AA~ which includes the \ion{Fe}{ii} 2600\AA~ resonance line. At day 4 a recombination wave occurs. }
\label{fig:EW1}
\end{figure}

\subsection{The He II line profiles variations}
\label{section_HeII}

In the early days after the eruption opacity effects may cause shadowing of the red-shifted emission originating in the receding part of the ejecta by the approaching blue-shifted part. 
The suppression of the positive velocities in the 2016 spectrum from day 3.95 shown in Fig.~\ref{fig:imacs2016_velocity} suggest that this mechanism could be operating. However, there is another explanation: 
The 1990 IUE spectra, see Fig.~\ref{fig:iue_swp}, show mostly symmetric line profiles in nearly all lines except in the He~II 1640\,\AA~ line. 
We illustrate the evolution of the He~II 1640 line profile, see Fig. \ref{fig:HeII1640}, by  normalizing to the light curve. 
The He II 1640~\AA\  profile is markedly different at orbital phase $\sim$0.7 (days 1.92, 1.97) compared with phase 0.1 and 0.3 (days 2.42 and 2.67, respectively). The spectra at phase 0.7 have more emission on the red wing centred on about 1660~\AA\  and less on the blue wing than the spectra at phase 0.1 and 0.3. 
Similarly, the day 3.86 (phase 0.21--0.26) spectrum matches those of day 2.42 and 2.76 (phase 0.1, and 0.3) in peaking in the blue, but then the day 5.77 (phase 0.68--0.79) spectrum is more like the day 1.92 and 1.97 (phase $\sim$0.7). The location of the peak is related to having a similar orbital phase.  Such line profile changes are limited to the He\,II\,1640 line, even though the other lines of ions with high ionisation energy (N\,V~1242, Si/O\, IV~1402/1406, C\,IV 1550), do not show much change to the overall profile prior to day 6, just to the flux.  Hence, the asymmetric profile of He~II lines in the 2016, day 3.95 (phase 0.23) spectrum may also be due to a variable component.

We conclude that in the 1990 eruption during day 2--9 the variable component of the He II 1640 emission line is tied to the orbital motion of the binary system, while the emission in H, He~I, and the N~III, C~III, C~IV, and Si~IV is not; those are formed in the ejecta. After day 8 a narrow component is a new addition to the He~II profile; we discuss narrow profiles in Section~\ref{sec:narrow}. The variability by orbital phase in He~II is possibly related to the accretion reestablishing which we discuss in Section~\ref{sec:accretion}.
This might mean that the donor star is He rich \citep{shore}, or that part of the WD atmosphere is ejected during the eruption but remains bound to the system.

\subsection{The recombination wave in the Fe~II UV 1 feature at 2600\AA}

\label{section_L2outflow}

In the UVOT spectra (Fig.~\ref{fig:uvot_spectra}) we see on days 4--6,  evidence for  \ion{Fe}{ii} absorption with features that extend over the whole range of 2300--3100\AA.  Inspection of the 1990 LWP IUE spectra (Fig.~\ref{fig:iue_swp} ) also showed the \ion{Fe}{ii} 2600\,\AA~ line.  To better understand how this absorption is formed we determined the absorption fraction under the continuum of the reddening corrected spectra in the 2550--2625\AA~ band. The continuum for each spectrum was fitted with a fixed slope ensuring that the continuum was applied in a consistent way, and checking that the 
normalised spectra matched in the spectral regions least affected by the absorption features or emission lines. The errors were determined by varying the normalisation.  The measured value is shown in Fig.~\ref{fig:EW1} as a function of time after eruption. 

The sudden strengthening of the \ion{Fe}{ii} absorption on day 4, and its subsequent decline to the previous level on day 9 is possibly due to a recombination wave. Recombination waves can be due to changes in opacity, temperature and ionising flux.  Whilst the peak occurred around phase 0.5, earlier measurements did not match the later rise, so it is thought to be unlikely to be due to an orbital effect.  After day 4 the ionising flux from the SSS phase starts to increase and cause the \ion{Fe}{ii} absorption decrease. 

During day 5.2 the overall UV flux in the spectra from 1700--3000\,\AA~ is lower and the  $uvw1$  photometry of day~5.22 is also fainter. This is probably due to the recombination wave as well.

\begin{figure}
\includegraphics[width=\columnwidth]{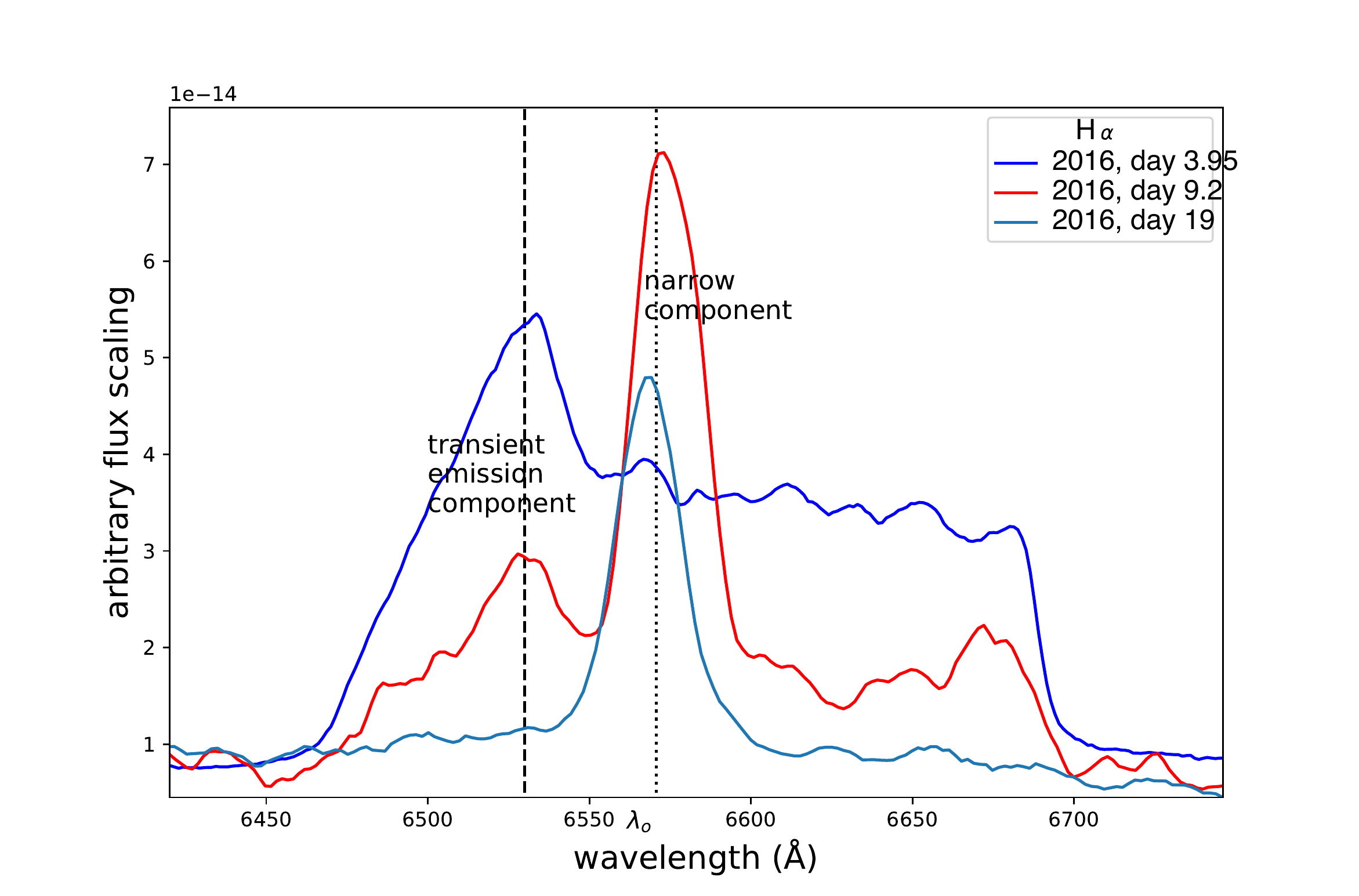}
\caption{Sample line profile changes in H$\alpha$ during the 2016 eruption (for the full spectra see Fig.~\ref{fig:optical_spectra}). Initially, the line is very asymmetric and skewed with a transient blue peak which is seen around day 4, and weakly day 8 (not shown). No narrow central peak is found until day 8; it is well-defined day 9.2 and it is still present on day 19. On day 68 it is no longer prominent. The spectra are plotted with observed wavelength, and  the dotted line is the central wavelength of H$\alpha$ corrected for the systemic velocity of the LMC of 278 km/s. 
}
\label{fig:Halpha2016}
\end{figure}

\subsection{Narrow line profile components}
\label{sec:narrow}
In our analysis we use spectra of the 1990 and 2016 outbursts.  
In Figures~\ref{fig:optical_spectra}, and \ref{fig:Halpha2016} a centred narrow line component can be seen in the H and He lines in 1990 and 2016 spectra after day 8 of the eruption and lasting at least till day 21.


A comparison of the 2016 FLOYDS spectrum (day 9.2, phase 0.63), of the 2016 spectrum from Mirranook Armidale (day 8.0, phase 0.73), of the first spectrum from \citet{sekiguchi} of the 1990 eruption obtained on 1990 Feb. 22 (day 8.73), and the 1990 CTIO spectrum of day 9.0 (phase 0.3), Fig.~\ref{fig:optical_spectra}, shows that none of of them exhibits any differences in the narrow component with orbital phase. 
  
Since the Balmer decrements show that the narrow component is due to recombination, we can expect a similar behaviour both in H and He lines and use that to determine when the narrow component appeared. 
Between day 4 and day 8,  a 1990 IUE spectrum was taken on day 5.8, which can be compared to the IUE day 8.9 spectrum. 
There is a central peak present in the day 8.9 spectrum in the He\,II 1640 line that is not seen in the day 5.8 spectrum, suggesting that the narrow line developed between day 6 and 8 (since we see it in the day 8 optical spectrum). 

Narrow components were also seen in other short period recurrent novae. A well-observed sequence can be seen in Nova LMC 2009a where the narrow component is seen to appear around day 10 and disappear when the SSS phase ends and thus the X-ray luminosity drops \citep{2016ApJ...818..145B}. 

\cite{williams1981} describe IUE observations of the 1979 outburst of nova U Sco where He\,II~1640 developed a narrow component between day 6 and 8 after discovery. 
\cite{sekiguchi1988} display a sequence of spectra for the 1987 outburst which shows no narrow peak on day 4, but it is present on day 9 and 19 after eruption. 
The well-observed 2010 eruption shows that on day 5 there was no central narrow component \citep{kafka2011,mason2012}, while it is present on the day 5.8 spectrum  and disappeared between days 33 and 41 \citep{anapuma2013}, comparable to the end of the SSS phase around day 40 \citep{orio2013}.

RN V394 CrA, a fast nova showing similarities to U~Sco, had two outbursts, one in 1947 \citep{duerbeck1988}, and one in 1987 \citep{liller1987}. Spectra taken 1987 Aug. 3 and 4, 5 and 6 days after the eruption, show only a broad flat line profile in the H\,I and He\,II lines, while by 1987 Aug. 13, day 11 and later, they show a narrow line profile component in these lines which appears diminished by Sep. 15. A spectrum taken 1987 Oct. 4 shows that the narrow component has all but disappeared \citep{sekigushi1989, starrfield1988}.

The narrow lines are thus found in similar RNe systems, and are characterised by the initial absence until several days after the eruption, and disappearance at the time the SSS emission also drops off. This suggests that the narrow emission is powered by the high-energy emission from the WD, just like the soft X-rays are. 

 Several mechanisms have been proposed to explain these narrow lines centered on the line wavelength and on top of a broad pedestal. 
Almost certainly the narrow component of the line is formed in a different region from the broad pedestal which is due to the ejecta. 
We explore the possibility here, that the narrow component is due to reionisation of trailing clumps of cool matter from the previous eruption at distances of several $10^{16}$~cm. 
The time delay between the eruption and the appearance of the narrow lines is  due to the light travel time to reionise the distant shell, assuming the UV/EUV radiation from the photosphere is sufficient even before the start of the SSS phase. 
The observed line width (FWHM) of $\approx 1100\,{\rm km}\,{\rm s}^{-1}$ will be due to a low gas temperature and velocity of the clumps, and is consistent with the distance reached in about seven years by the slower ejecta. 
The recombination time needs to be less than about a day which requires densities of $>10^7 {\rm cm}^{-3}$ and implies a filling factor of $<10^{-6}$.
Estimated filling factors in the days after the eruption of 0.01--0.001 scale after 7 years of expansion to $<10^{-6}$ provided the clumps do not grow faster than by thermal expansion.

We cannot reliably detect variations in the narrow component with orbital phase in N\ LMC 1968. 
Reionisation of clumpy ejecta from a previous eruption provides a possible explanation for the narrow line profiles.

\section{Observations in a model context}
\label{sec:model}
\subsection{A comprehensive model}


Late time spectral observations show that nova ejecta have a bipolar shape \citep{hutchings1972,mustel1970,solf1983,gill1999,gill2000,harman2003,ribeiro2009,shore2012, shore2013_tpyx, shore2013_mon, shore2016_v339del}, and we will assume this to be so for this nova. 
The visible brightness decay time scale t$_3$, is shorter for smaller mass ejected and higher velocity since the ejecta become transparent faster.
In the RNe with a late-type secondary the SSS X-ray emission onset is delayed from the eruption itself by a period of at least a few days \citep{Schwarz2011}. The SSS emission is likely the surface emission of the WD with nuclear burning continuing after the initial eruption for a period of weeks to months until the all the hydrogen has been burned. The upper atmospheric temperature of the WD is typically $<$1MK.  

The accretion disk may have been disrupted in the initial eruption  \citep[for U Sco see][]{drakeorlando}, though the ejected mass is low, reducing the potential for disruption. For example, in M31 N 2008-12a the accretion disk is thought to survive \citep{henze2018}. In addition to the eruption, the high luminosity of the WD will heat the atmosphere of the secondary. For a secondary that fills its Roche lobe this leads to matter filling the inner WD Roche lobe from both the accretion disk and by overflow from the heated  atmosphere.   Eventually that matter will undergo a pancake-like instability and reform the accretion disk. 
At the same time, over possibly a long period, hydrogen is converted to helium in the layers still bound to the WD, and not lost. 

Thus, in the higher mass WDs ejecta become transparent faster, reducing the time for the density of the ejecta to become low enough to observe the WD and inner system. 
N\ LMC 1968 has a very small $t_3$, smaller than the Galactic nova U~Sco which also has an evolved companion, similar spectra, and SSS phase, and a comparable orbital period of 1.23d \citep{schaefer1995}.
In N LMC 1968 the suggestion of the changing eclipse profile, the early occurrence of eclipses, and the behaviour of He~II (see Section~\ref{section_HeII}) are all consistent with a model where debris filling up the inner WD Roche lobe  initially block the WD photosphere from view, though that does not rule out that the bipolar ejecta also cross the line of sight. 
We'll consider the consequences of that in the following subsections. 

\subsection{Inclination of the orbit}
\label{inclination}

It is likely that we are seeing the system at an inclination angle close to the orbital plane because we have seen eclipses. 
\citet{Ness2013} estimate an inclination angle of $>76^{\rm o}$  is to be expected for an SSe type system (which shows emission lines).
Our interpretation of the UV flux variations suggests that we view close across the rim of the accretion disk, which would imply a system inclination in the 68--76$^{\rm o}$ range:
\cite{warpeddisk} show that for a low $\alpha$, thicker disk, the disk can  warp and rotate rigidly, which could fit with the observed changes to the UV emission. 

The SSS onset could be expected to happen sooner if the line of sight were not intersecting the bipolar ejecta, i.e., at high inclination. The presence of an accretion disk does not preclude scattering of soft X-rays in a halo above the disk. However, in such a situation the debris left in the inner Roche lobe in the days after the eruption could effectively block the X-rays and delay the SSS onset. 

\subsection{The accretion disk and precursor}
\label{sec:accretion}
It is unknown what happens to the accretion disk during and after the nova explosion, but N LMC 1968 may provide some insight.  
The main questions relating to the accretion disk are: (1) was the accretion disk destroyed in the explosion, (2) does the Roche lobe overflow (RLOF) from the secondary change, and (3) how long until the accretion disk is restored as a steady element in the system? 
The hydrodynamic models from \citet{drakeorlando} show that in the eruption of U~Sco, a very similar system, the accretion disk would not survive the initial blast which seems to answer point (1). However, we do not have the data to support that. The accretion disk may very well have survived the blast; perhaps in a state where the vertical structure was no longer in hydrostatic equilibrium.
As to point (2), our observations show evidence of matter bound to the system in the UV variability, in the changes to the He~II profiles as early as day 2, and RLOF is possibly a factor in the delay of the onset of the SSS.
As to point (3): if indeed the 'debris' is sufficiently spread out throughout the inner Roche Lobe and neighboring regions, that matter, whilst blocking the X-rays from the WD atmosphere, would need to collapse to the orbital plane to reform the initial accretion disk. Taking a look at the observations they seem to indicate the process to take 6 days, at which time the SSS starts and the gradient of the light curve steepens, in accordance to a change in the optical and UV source. 

The physical process can be modelled simply by assuming that after the eruption we start reformation of the accretion disk with a messy atmosphere filling the Roche lobe around the WD. 
Whether this atmosphere consists of large blobs moving under gravity nearly like solid bodies or is fully turbulent, the momentum in vertical motions will dissipate with each crossing of the orbital plane, since atmospheric elements with opposite momentum will interact. 
We can identify the spectral signature of this turbulent material with the UV-optical spectral energy distribution (SED). The variable line components seen in He~II are taken as an indication of the blob  velocities of order 3000~km~s$^{-1}$ (the He~II FWZI is 10,000km~s$^{-1}$). 
With the disk formation and clearing out of the WD Roche lobe within 6 days  the disk is not necessarily stationary, since the mass inflow may still evolve either from the secondary or from enhanced inflow  from the heated atmosphere of the secondary. 
Since orbits closer to the WD mean higher velocities, dissipation of energy and relaxation times will, similarly to an accretion disk, be larger near the WD. 
This scenario provides an update  to the simple picture of the formation of an accretion disk by \citet[][]{verbunt1982} for novae.


\subsection{H burned to He during the SSS-phase}
\label{sec:he_created}
During the SSS phase  steady nuclear burning takes place on the surface of the WD. The burning is eventually quenched due to a drop in the temperature and pressure, which in turn happens when the hydrogen that fuels the luminosity runs low. We can assume that the steady burning occurred from the time of the explosive ejection of material until the time of turnoff. The SSS luminosity is thus a way to measure how much He is added to the WD during that period, regardless of the source of the H being pre- or post-eruption. The difference between He ejected and added due to burning during the SSS phase gives an indication of the rate of growth of the WD mass. 

A blackbody fit to the X-ray spectrum shows that during the SSS phase the luminosity of the surface is $1.1 \times 10^{37}$~erg~s$^{-1}$ (Section \ref{sec:uv_xray_ratio}, Fig.~\ref{fig_xrt_spectra_and_BBfit}). 
The downturn in the light curve starts at day 30, so in 30 days the energy produced was $2.8 \times 10^{43}$~erg. 
Assuming one gram of H converted to $^4$He produces $6.40\times 10^{+18}$~erg, the 
total He mass created is  M(He) = $4.4\times 10^{24}g =2.2\times 10^{-9}M_\odot$.
If the accreting matter were high in He, the estimated growth in WD mass is a lower limit.

In N LMC 1968 the He abundance is very high \citep{shore}. This may be due to He rich ejecta or because the secondary has lost most of its hydrogen.  In our proposed messy atmosphere the variable He~II line component would be in the inner Roche lobe in the early days after the eruption but it is unclear how it got there.

\begin{table}
\caption{Summary of the parameters for N\ LMC 1968 }
\label{table_results}

\begin{tabular}{@{}lcccrrrr}
\hline
property & value & unit & Section \\
\hline
2016 eruption JD& $2457408.709\pm 0.8$ & & \ref{sec:nova2016}\\
adopted distance LMC & 50 & kpc & \ref{sec:nova1990b} \\
minimum mag $<V_{min}>$ & $19.70       $ & mag&  \\
maximum mag $V_{max}$ & $12.3 \pm 0.5$ & mag&  \\
decay time 2 mag $t_2$     & $4.6 \pm 0.5$ & d & \ref{sec:lc}\\
decay time 3 mag $t_3$     & $7 \pm 1$      & d & \ref{sec:lc}\\
epoch binary & $2455058.323\pm 0.090$ & HJD & \ref{sec:orbit}\\
period & $1.26433 \pm 0.000019$& d & \ref{sec:orbit}\\
period change $\Delta P/P$ & < 0.003 & & \ref{sec:orbit} \\
reddening E(B-V) & 0.07 & mag & \ref{sec:reddening}\\
interstellar $N_H$ & $1.8 \times 10^{21}$ & $cm^{-2}$ & \ref{sec:xrtspectra} \\
SSS emission phase & 6-57 & d & \ref{sec:xrtlightcurve}\\
SSS typical luminosity & $1.1 \times 10^{37}$& erg/s&\ref{sec:xrtspectra}\\
Kinetic energy & $\approx 10^{38}$& erg/s&\ref{sec:uv_xray_ratio}\\
SSS typical BB temp.(kT) & 100 & eV & \ref{sec:xrtspectra} \\
narrow component FWZI  & $1.6\times 10^3$ & km s$^{-1}$ & \ref{sec:optical_spectroscopy}\\
FWZI broad component & 10$^4$ & km s$^{-1}$ &\ref{sec:optical_spectroscopy}\\
Sp. type secondary   & unknown   &             & \ref{sec:secondary}\\
Ejecta Velocity & 5000 & km s$^{-1}$ &\ref{sec:optical_spectroscopy} \\
WD mass & $> 1.3 $ & $M_\odot$ & \ref{sec:mass}\\ 
depth eclipse & 0.6 & mag & \ref{sec:orbit} \\
duration eclipse & 0.05 & phase & \ref{sec:orbit}\\
ejected mass & $ \approx 10^{-7}$ & $M_\odot$ & \ref{sec:massejected}\\
H converted to $^4$He & $2.2 \times 10^{-9}$ & $M_\odot$ & \ref{sec:he_created} \\
system inclination & $68-76$ & degrees & \ref{inclination}\\
recurrence time &$6.2 \pm 1.2$ & years &\ref{recurrencetime} \\
UV-optical L(1d) & $4\times 10^{37}$ & erg s$^{-1}$  & \ref{sec:uv_xray_ratio} \\
peak L$_X$(14-30d) & $1.1 \times 10^{37}$ &erg s$^{-1}$ & \ref{sec:uv_xray_ratio} \\
\hline
\end{tabular}
\end{table}


\subsection{Estimate of the ejected mass}
\label{sec:massejected}

The quiescent luminosity is thought to be  dominated by the accretion disk luminosity L$_{acc}$. Taking the photometry from around day 70 we can construct an SED (1900--9000\AA). 
We estimate L$_{acc} \ge 4\times 10^{34} {\rm erg s}^{-1}$ where the lower limit is because the dereddened spectrum has an unaccounted component which is still rising in the UV. This translates to an accreted mass of at least M$_{acc} \ge 3.5 \times 10^{-8} {\rm M}_\odot$ using \citet{osborne2011}. 

Under an assumption that the bipolar ejecta cross our line of sight, and there is no debris left in the inner Roche lobes, we can use  the start time of the SSS phase to derive an estimate of the ejected mass along the line of sight, since that happens when the  \NH\, column reaches a value where the soft X-rays become transparent while we also know at what time the SSS rise occurs.
Multiplying that time with the measured ejecta velocity then gives the distance to use with \NH.  
On day 7.2, \NH = $4 \times 10^{21}$ cm$^{-2}$, while V$_{ej}$ = 5000\,km s$^{-1}$.
Of course, for extrapolating that to the total ejected mass we will have to make some estimate of the geometry of the ejecta. 
Since the ejecta are bipolar, the ejecta covers a solid angle $\Omega$, and 
we derive an ejecta mass of $\approx 3\times 10^{-7}(\Omega/(4\pi)$ M$_\odot$. 
Assuming the accreted mass is 2-3 times the lower limit above and a solid angle of $\approx\pi$, the numbers for accreted and ejected mass are compatible with M$_{ej} \approx 10^{-7}  {\rm M}_\odot$, but do not allow a determination of net mass loss or growth for the WD.

\subsection{The UV and X-ray luminosities}
\label{sec:uv_xray_ratio}		

The peak UV luminosity (1200--3300\AA) of N LMC 1968 in the 1990 outburst was computed by \citet{shore} to exceed the Eddington luminosity ($1.6 \times 10^{38}$ erg s$^{-1}$ for a 1.3 M$_\odot$ WD), but with our adopted smaller distance, reddening and N$_{\mathrm H}$, the luminosity is nearly a factor 4 smaller: 
Using the SED derived from the normalisation of the light curve fit at day 1.0 (see Section~\ref{fig_uv_opt_ir_lc}) and fitting that with a blackbody, the peak luminosity of the UV-optical component is $L_{UV} \approx 4 \times 10^{37}$ erg s$^{-1}$. We also tried to fit the dereddened continuum of the combined day 2 to 4 spectra using the 1990 IUE as well as the 2016 day 2.9 UVOT and day 3.95 IMACS spectra, covering  1150--9000\AA, but a single blackbody fit is not possible. The UV rise suggests a temperature in excess of $\sim$25,000~K plus a cool component to deal with excess IR flux. 

After the initial ejection, the WD luminosity becomes thermalised by the optically thick ejecta, but once the ejecta have become optically thin, the luminosity comes out in the X-ray and EUV. 
We found in Section~\ref{sec:xrtspectra} that $L_{X-Bol} \approx 1.1\times 10^{37}$~erg s$^{-1}$. The UV-optical light during the SSS  is by then much less than near the peak: A black body fit for the UV-optical component at that time gives $L_{UV} = 2.4 \times 10^{35} $ erg s$^{-1}$.  

A review of the past eruptions suggest the rise time to peak is about a day. Using the estimate for the ejected mass and the expansion velocity, the kinetic energy rate imparted in the first day is around $3\times(\Omega/4\pi)\times10^{38}$~erg~s$^{-1}$, with $\Omega$ the solid angle of the ejecta. This suggests that the kinetic energy imparted the first day of the eruption is close to the Eddington luminosity, and dominates over the energy lost in radiation. 

\section{Conclusions}
\label{sec:results}

In Table~\ref{table_results} the main parameters which have been derived are collected. 

The recurrence time of N LMC 1968 is short and we expect another explosion to happen around April 2022 $\pm$ 1.2 years. In this study we collected and interpreted the known data from which we derived the ephemeris for the WD eclipse. Though the system shows eclipses, we do not have radial velocity measurements which would help determine the orbit and mass ratio of the binary.  

The radiant luminosity is found to be lower than Eddington for a He-atmosphere by an order of magnitude, unlike in many other novae. If the intrinsic luminosity of the nova is near Eddington, the lower observed value could be due to shadowing by the accretion disk, where only Thompson scattering of the X-rays in a halo above the orbital plane of the system is observed. However, the kinetic energy in the ejecta required an Eddington luminosity during the initial TNR.  This could mean that for the rapid recurrent systems the kinetic energy dominates over the observed radiative component. 

We also observe evidence that suggests that prior to day 7 (after the  eruption) there is a source of gravitationally bound material in the system, possibly due to increased mass flow from the secondary, disruption of the accretion disk, or part of the WD atmosphere that did not reach escape speed. 
This is seen as a variable He$^+$-rich matter component which shows  orbital periodicity day 2--7 after the 1990 eruption and could be located in the inner Roche lobe or just outside it.
The eventual collapse of this matter into the orbital plane could bring about the emergence of the WD soft X-ray emission. 
The rise of the SSS emission could alternatively be due to the decreasing column density in the expanding bipolar ejecta.  

A sudden increase in the Fe~II UV absorption on day 4 is seen as evidence for a reionisation wave happening in the ejecta. 

Linking the SSS luminosity to steady burning on the WD surface we arrive at an estimate of the mass of H converted to He after the eruption. This matter is adding to the WD mass, providing a link between SSS duration and growth of the WD. We also estimate the ejected mass and find that to be compatible with the accretion rate. The high SSS temperature suggests the WD mass is larger than 1.3 M$_\odot$.

In the UV we observe eclipses out to 320 days after the 2016 eruption which are shallow and broad while the emission is  variable,  suggesting that the  accretion disk is not stable or is perhaps warped. 

The current study has left many questions.  Some of those could be answered by future high resolution multispectral observations of the different phases of the eruption, and better characterisation of the orbital parameters. {\upd A comparison of the very similar Galactic nova U Sco to N LMC 1968, which has a similar orbital period $P_{orb} = 1.2305d$ \citep{schaefer2011} and recurrence time of $10 \pm 2$ years \citep{2010ApJS..187..275S} is planned for future work.}

\section*{acknowledgements}


{\small 
The OGLE project received funding by the National Science Center, Poland, 
under grant MAESTRO 2014/14/A/ST9/00121 to A.U. P.M. acknowledges support from the Foundation for Polish Science (Program START).
N.P.M.K., K.L.P., A.A.B., A.P.B. and J.P.O.\ acknowledge support from the U.K.\ Space Agency.
M.H.\ acknowledges the support of the Spanish Ministry of Economy and Competitiveness (MINECO) under the grant FDPI-2013-16933.
S.S. acknowledges partial support from NASA, HST, \& NSF grants to ASU.
Research in Novae at Stony Brook University is supported in part by NSF grant AST 1614113, with additional research support provided by the Stony Brook University. RDG was supported by NASA and the United States Air Force.
V.A.R.M.R. acknowledges financial support from the Funda\c{c}\~{a}o para a Ci\^encia e a Tecnologia (FCT) in the form of an exploratory project of reference IF/00498/2015, from the Center for Research \& Development in Mathematics and Applications (CIDMA) strategic project UID/MAT/04106/2019, and supported by Enabling Green E-science for the Square Kilometre Array Research Infrastructure (ENGAGE-SKA), POCI-01-0145-FEDER-022217, and PHOBOS, POCI-01-0145- FEDER-029932, funded by Programa Operacional Competitividade e Internacionaliza\c{c}\~ao (COMPETE 2020) and FCT, Portugal.
RA acknowledges financial support from DIDULS Regular PR\#17142 by Universidad de La Serena.
The Swift data were retrieved from the UK Swift Data centre.
IUE spectra were retrieved from the MAST archive.  
This work makes use of observations from the Las Cumbres Observatory  network.  
We acknowledge with thanks the variable star observations from the AAVSO International Database contributed by observers worldwide and used in this research.
We would like to acknowledge Elena Mason, and Bob Williams for discussions, Mike Shara, Kaz Sekiguchi, and David Buckey who tried to chase down old observations of this nova, and Patrick Godon for discussions on the accretion disk. 
We used HEASARC Ftools, IDL, scipy, matplotlib, and astropy.  
} 

\bibliographystyle{mnras}
\bibliography{nlmc} 

\appendix

\section{Appendix}

\begin{table*}
\begin{minipage}{150mm}
\caption{All known UV-optical spectra of N\ LMC 1968 ordered by day since respective outburst. }
\begin{tabular}{@{}rrrrrrrrrrl}
\label{tab:all_spectra}
mission or &year &midtime      &day   &orbital     &ID and notes\\      
instrument &eruption& JD           &since  &phase   &     \\             
   (1)     &     &    (d)           &eruption & (2) & \\
\hline
CTIO/Argus &1990 &2447936.6      &0.0   &0.19       &discovery spectrum (3)\\
IUE        &1990 &2447938.51149  &1.92  &0.70       &SWP38199 \\
IUE        &1990 &2447938.55107  &1.95  &0.73       &LWP17374 \\
IUE        &1990 &2447938.56480  &1.97  &0.75       &SWP38200 \\
IUE        &1990 &2447939.01385  &2.42  &0.10       &SWP38202 \\
IUE        &1990 &2447939.03438  &2.44  &0.12       &LWP17378  \\
IUE        &1990 &2447939.23461  &2.65  &0.27       &LWP17379 \\
IUE        &1990 &2447939.26223  &2.67  &0.30       &SWP38204 \\
IUE        &1990 &2447939.28532  &2.69  &0.31       &LWP17380 \\
UVOT UVG   &2016 &2457410.78200  &2.88  &0.64       &00045768005\\
UVOT UVG   &2016 &2457411.32973  &3.43  &0.07       &00045768006\\
IUE        &1990 &2447940.44478  &3.86  &0.21-0.26  &SWP38209   \\
IUE        &1990 &2447940.47694  &3.89  &0.23-0.29  &LWP17390 \\
UVOT UVG   &2016 &2457411.77783  &3.88  &0.4260     &00045768007\\ 
LCO IMACS  &2016 &2457411.5351  &3.945 &0.23       &fits file    \\    
UVOT UVG   &2016 &2457411.91122  &4.01  &0.5315     &00045768008\\
UVOT UVG   &2016 &2457412.97519  &5.07  &0.3730     &00045768009\\
UVOT UVG   &2016 &2457413.12185  &5.22  &0.4890     &00045768010\\
IUE        &1990 &2447942.34466  &5.77  &0.68-0.79  &SWP38214 \\
UVOT UVG   &2016 &2457414.04720  &6.14  &0.2209     &00034302002\\
UVOT UVG   &2016 &2457415.11269  &7.21  &0.0636     &00034302004\\
SAAO       &1990 &2447944.31     &7.71  &0.29       &not recoverable \\
Mirranook Armidale&2016 &2457415.9501 &8.05 &0.73 &fits file \\
UVOT UVG   &2016 &2457416.03941  &8.13  &0.7966     &00034302006\\
SAAO       &1990 &2447945.33     &8.73  &0.10       &published\\
UVOT UVG   &2016 &2457416.70604  &8.80  &0.3239     &00034302008\\
IUE        &1990 &2447945.50490  &8.93  &0.71-0.82  &SWP38229 \\
CTIO       &1990 &2447945.6029   &9.00  &0.31       &fits file\\
FTS/FLOYDS &2016 &2457417.1      &9.2   &0.63       &fits file\\
IUE        &1990 &2447946.25125  &9.69  &0.74-0.91  &SWP38231 \\
UVOT VG    &2016 &2457417.63602  &9.73  &0.5940     &00034302010\\
UVOT VG    &2016 &2457418.03603 &10.13  &0.3758     &00034302012\\
SAAO       &1990 &2447948.29    &11.69  &0.44       &not recoverable \\
SAAO       &1990 &2447952.36    &15.76  &0.66       &published\\
IUE        &1990 &2447952.88444 &16.35  &0.95-0.20  &SWP38284 \\
LCO duPont &2016 &2457427.6486  &19.73  &0.98       & fits file\\
LCO duPont &2016 &2457428.6344  &20.72  &0.76       & fits file\\
LCO duPont &2016 &2457429.5827  &21.67  &0.51       & fits file\\
IUE        &1990 &2447970.87319 &34.35  &0.15-0.45  &SWP38394 \\
IUE        &1990 &2447976.81065 &40.35  &0.72-0.27  &SWP38439 \\
IUE        &1990 &2447977.81134 &41.35  &0.73-0.28  &LWP17625 \\
LCO IMACS &2016 &2457476.5447  & 68.64 &  0.65& fits file\\
LCO IMACS &2016 &2457476.5690  & 68.67 &  0.67& fits file\\
\hline

\end{tabular}
\begin{tabular}{@{}lrrrrrrrrrl}

$^1$ Spectral ranges per instrument:  \\
IUE SWP is 1150--2000A; IUE LWP is 1800--3300A, UVOT UVGRISM is 1700--5000A; \\
UVOT VGRISM is 2900--6600A; CTIO is 3500--7700A; Magellan/IMACS is 4000--9000A;\\
SAAO is 3500--7200A; FTS/Floyds 3300--11000A.  \\
$^2$ for the 1990 eruption the orbital phase uncertainty is 0.15 and based on Eq. \ref{eq1} \\
The range in phase is due to long observations, e.g. for IUE, otherwise <=0.02.\\
$^3$ This has been lost, unfortunately. Personal communication with Mike Shara and Bob Williams.
\end{tabular}

\end{minipage}  
\end{table*}

\begin{table*}
\begin{minipage}{150mm}
\caption{Swift UVOT Grism exposures. }
\label{table_grism_obs}
\begin{tabular}{@{}rrrrrrrrrrl}
\hline
Mid-time$^1$&Day$^2$ &Orbital$^1$&{\it Swift} OBSID &Roll & Anchor$^5$  &Exposure & UV/V &$uvw1^3$&Scale$^4$&  \\
JD(+2450000)&        &phase   &             & deg &(X,Y)pix &time (s) &grism &       &factor \\
\hline
7410.78200& 2.88&.6383 &00045768005& 219.1& 1046, 875& 285.3 &UV&13.642&1.00&\\
7411.32973& 3.43&.0716 &00045768006& 223.0& 1235, 925& 405.0 &UV&13.723&1.08&\\
7411.77783& 3.88&.4260 &00045768007& 216.0& 1198,1034& 282.3 &UV&13.697&1.05&\\
7411.91122& 4.01&.5315 &00045768008& 220.0& 1263, 945& 285.9 &UV&13.694&1.05&\\
7412.97519& 5.07&.3730 &00045768009& 217.0& 1270, 834& 294.3 &UV&14.363&1.95&\\
7413.12185& 5.22&.4890 &00045768010& 221.1& 1106, 886& 398.8 &UV&14.624&2.48&\\ 
7414.04720& 6.14&.2209 &00034302002& 237.0& 1130,1613& 892.5 &UV&15.614&6.16&\\ 
7415.11269& 7.21&.0636 &00034302004& 222.0&  972,1620& 946.6 &UV&15.822&7.46&\\
7416.03941& 8.13&.7966 &00034302006& 223.0& 1034,1590& 999.7 &UV&15.884&7.90&\\
7416.70604& 8.80&.3239 &00034302008& 223.0&  952,1636& 892.5 &UV&16.156&10.15&\\
7417.63602& 9.73&.5940 &00034302010& 232.0& 1114,1662& 892.5 &V &16.586&15.08&\\
7418.03603&10.13&.3758 &00034302012& 232.0& 1101,1638& 892.5 &V &16.675&16.37&\\
\hline
\end{tabular}
$^1$ mid-time of exposure  \\
$^2$ days since estimated time of eruption JD\,2457407.9 \\
$^3$ uvw1 magnitude interpolated from a spline fit to the light curve \\
$^4$ flux scale factor derived from uvw1 light curve\\
$^5$ the anchor is defined by the position in the first order spectrum at 260\,nm (UV) or 420\,nm (V)\\  
\end{minipage}  
\end{table*}

\begin{table}[]
\caption{Photometry from 10 days before to 50 days after the 2016 eruption of N\ LMC 1968$^a$}
\label{tab:photometry}
\begin{tabular}{@{}lrrrrrrrrrl}
\hline
Date$^a$ & mag & mag&Orbital$^c$& instrument  &phot.&  \\
(MJD)    &     & error    &phase      & and filter   & system &  \\
\hline
7399.24670 & 19.297 & 0.042 & 0.91014 & OGLE I & Vega \\
7404.23630 & 19.422 & 0.042 & 0.85658 & OGLE I & Vega \\
7408.20942 & 99.999 & 9.999 & 0.99906 & OGLE I & Vega \\
7410.04082 & 12.820 & 0.005 & 0.44758 & Andicam I & Vega \\
7410.04157 & 13.298 & 0.004 & 0.44817 & Andicam B & Vega \\
7410.05324 & 13.556 & 0.008 & 0.45740 & Andicam V & Vega \\
7410.05391 & 12.605 & 0.005 & 0.45793 & Andicam R & Vega \\
7410.17573 & 12.774 & 0.056 & 0.55428 & ANS & Vega \\
7410.17573 & 13.211 & 0.032 & 0.55428 & ANS & Vega \\
7410.17573 & 13.228 & 0.028 & 0.55428 & ANS & Vega \\
7410.21081 & 12.930 & 0.003 & 0.58202 & OGLE I & Vega \\
7410.27711 & 13.643 & 0.023 & 0.63446 & UVOT UVW1 & AB \\
7410.82572 & 13.710 & 0.027 & 0.06838 & UVOT UVW1 & AB \\
7411.17503 & 13.110 & 0.029 & 0.34466 & ANS & Vega \\
7411.17503 & 13.573 & 0.013 & 0.34466 & ANS & Vega \\
7411.17503 & 13.628 & 0.022 & 0.34466 & ANS & Vega \\
7411.27294 & 13.647 & 0.023 & 0.42210 & UVOT UVW1 & AB \\
7411.40628 & 13.793 & 0.024 & 0.52756 & UVOT UVW1 & AB \\
7412.06793 & 13.868 & 0.012 & 0.05088 & Andicam I & Vega \\
7412.06864 & 14.110 & 0.007 & 0.05145 & Andicam B & Vega \\
7412.06939 & 14.316 & 0.016 & 0.05204 & Andicam V & Vega \\
7412.07005 & 13.825 & 0.011 & 0.05256 & Andicam R & Vega \\
7412.08291 & 13.768 & 0.003 & 0.06273 & OGLE I & Vega \\
7412.17229 & 13.808 & 0.040 & 0.13342 & ANS & Vega \\
7412.17229 & 14.030 & 0.023 & 0.13342 & ANS & Vega \\
7412.17229 & 14.117 & 0.019 & 0.13342 & ANS & Vega \\
7412.24426 & 13.774 & 0.012 & 0.19035 & Andicam I & Vega \\
7412.25706 & 14.180 & 0.015 & 0.20047 & Andicam V & Vega \\
7412.25797 & 13.620 & 0.012 & 0.20119 & Andicam R & Vega \\
7412.47017 & 14.052 & 0.024 & 0.36902 & UVOT UVW1 & AB \\
7412.61739 & 14.902 & 0.031 & 0.48547 & UVOT UVW1 & AB \\
7413.16938 & 14.257 & 0.058 & 0.92206 & ANS & Vega \\
7413.16938 & 14.531 & 0.034 & 0.92206 & ANS & Vega \\
7413.16938 & 14.597 & 0.030 & 0.92206 & ANS & Vega \\
7413.17357 & 14.234 & 0.003 & 0.92537 & OGLE I & Vega \\
7413.53753 & 14.515 & 0.024 & 0.21323 & UVOT UVW2 & AB \\
7413.55142 & 14.574 & 0.030 & 0.22422 & UVOT UVW2 & AB \\
7414.13745 & 14.791 & 0.048 & 0.68774 & Andicam I & Vega \\
7414.14558 & 15.311 & 0.013 & 0.69417 & Andicam B & Vega \\
7414.14966 & 14.998 & 0.009 & 0.69739 & Andicam V & Vega \\
7414.15374 & 14.715 & 0.014 & 0.70062 & Andicam R & Vega \\
7414.16670 & 14.390 & 0.072 & 0.71087 & ANS & Vega \\
7414.16670 & 14.586 & 0.028 & 0.71087 & ANS & Vega \\
7414.16670 & 14.839 & 0.071 & 0.71087 & ANS & Vega \\
7414.26437 & 14.501 & 0.003 & 0.78812 & OGLE I & Vega \\
7414.60489 & 15.596 & 0.039 & 0.05745 & UVOT UVM2 & AB \\
7414.61739 & 15.732 & 0.036 & 0.06733 & UVOT UVW2 & AB \\
7415.15130 & 15.604 & 0.008 & 0.48962 & Andicam I & Vega \\
7415.15305 & 15.983 & 0.006 & 0.49101 & Andicam B & Vega \\
7415.15484 & 16.013 & 0.010 & 0.49242 & Andicam V & Vega \\
7415.15605 & 15.573 & 0.008 & 0.49338 & Andicam R & Vega \\
7415.16300 & 15.396 & 0.003 & 0.49888 & OGLE I & Vega \\
7415.52989 & 15.873 & 0.031 & 0.78906 & UVOT UVW1 & AB \\
7415.54447 & 16.009 & 0.039 & 0.80059 & UVOT UVW2 & AB \\
7416.15497 & 16.253 & 0.014 & 0.28346 & Andicam I & Vega \\
7416.15677 & 16.565 & 0.008 & 0.28488 & Andicam B & Vega \\
7416.15852 & 16.591 & 0.014 & 0.28627 & Andicam V & Vega \\
7416.15977 & 16.193 & 0.013 & 0.28726 & Andicam R & Vega \\
7416.16446 & 15.929 & 0.132 & 0.29096 & ANS & Vega \\
7416.16446 & 16.228 & 0.054 & 0.29096 & ANS & Vega \\
7416.16446 & 16.260 & 0.071 & 0.29096 & ANS & Vega \\
\hline
\end{tabular}
\end{table}
\begin{table}[]
\contcaption{}
\begin{tabular}{@{}lrrrrrrrrrl}
\hline
Date$^a$ & mag & mag&Orbital$^c$& instrument  &phot.&  \\
(MJD)    &     & error    &phase      & and filter   & system &  \\
\hline
7416.18044 & 16.216 & 0.004 & 0.30360 & OGLE I & Vega \\
7416.21044 & 16.167 & 0.064 & 0.32733 & UVOT UVW2 & AB \\
7417.03829 & 16.251 & 0.043 & 0.98211 & ANS & Vega \\
7417.03829 & 16.780 & 0.019 & 0.98211 & ANS & Vega \\
7417.03829 & 17.075 & 0.028 & 0.98211 & ANS & Vega \\
7417.13390 & 16.951 & 0.016 & 0.05773 & Andicam I & Vega \\
7417.13569 & 17.117 & 0.009 & 0.05915 & Andicam B & Vega \\
7417.13744 & 17.225 & 0.017 & 0.06053 & Andicam V & Vega \\
7417.13869 & 16.809 & 0.015 & 0.06152 & Andicam R & Vega \\
7417.14031 & 16.660 & 0.053 & 0.06279 & UVOT UVW1 & AB \\
7417.52781 & 16.548 & 0.042 & 0.36928 & UVOT UVW1 & AB \\
7417.54031 & 16.738 & 0.055 & 0.37917 & UVOT UVW1 & AB \\
7418.08238 & 15.981 & 0.034 & 0.80791 & ANS & Vega \\
7418.08238 & 16.408 & 0.015 & 0.80791 & ANS & Vega \\
7418.08238 & 16.571 & 0.008 & 0.80791 & ANS & Vega \\
7418.12675 & 16.676 & 0.014 & 0.84301 & Andicam I & Vega \\
7418.12850 & 16.904 & 0.007 & 0.84439 & Andicam B & Vega \\
7418.13025 & 16.965 & 0.013 & 0.84578 & Andicam V & Vega \\
7418.13150 & 16.719 & 0.013 & 0.84677 & Andicam R & Vega \\
7418.45558 & 16.740 & 0.097 & 0.10309 & UVOT UVW1 & AB \\
7419.10888 & 16.940 & 0.015 & 0.61981 & Andicam I & Vega \\
7419.11063 & 17.161 & 0.008 & 0.62119 & Andicam B & Vega \\
7419.11238 & 17.141 & 0.015 & 0.62258 & Andicam V & Vega \\
7419.11363 & 16.940 & 0.015 & 0.62357 & Andicam R & Vega \\
7419.15646 & -1.000 & -1.000 & 0.65744 & ANS & Vega \\
7419.15646 & 16.570 & 0.018 & 0.65744 & ANS & Vega \\
7419.15646 & 16.806 & 0.016 & 0.65744 & ANS & Vega \\
7420.04870 & -1.000 & -1.000 & 0.36314 & ANS & Vega \\
7420.04870 & 16.610 & 0.016 & 0.36314 & ANS & Vega \\
7420.04870 & 16.770 & 0.038 & 0.36314 & ANS & Vega \\
7420.11907 & 16.882 & 0.017 & 0.41879 & Andicam I & Vega \\
7420.12086 & 17.151 & 0.008 & 0.42021 & Andicam B & Vega \\
7420.12261 & 17.152 & 0.015 & 0.42159 & Andicam V & Vega \\
7420.12386 & 16.869 & 0.016 & 0.42258 & Andicam R & Vega \\
7420.25142 & 16.710 & 0.026 & 0.52348 & UVOT UVW1 & AB \\
7420.65281 & 16.884 & 0.031 & 0.84095 & UVOT UVW1 & AB \\
7421.03317 & 16.153 & 0.045 & 0.14179 & ANS & Vega \\
7421.03317 & 16.739 & 0.024 & 0.14179 & ANS & Vega \\
7421.03317 & 16.900 & 0.032 & 0.14179 & ANS & Vega \\
7421.15504 & 16.899 & 0.016 & 0.23818 & Andicam I & Vega \\
7421.15679 & 17.028 & 0.007 & 0.23956 & Andicam B & Vega \\
7421.15854 & 17.035 & 0.014 & 0.24095 & Andicam V & Vega \\
7421.15979 & 16.885 & 0.015 & 0.24194 & Andicam R & Vega \\
7421.51739 & 16.646 & 0.031 & 0.52478 & UVOT UVW1 & AB \\
7421.58406 & 16.601 & 0.030 & 0.57751 & UVOT UVW1 & AB \\
7422.04978 & 16.396 & 0.043 & 0.94586 & ANS & Vega \\
7422.04978 & 16.812 & 0.039 & 0.94586 & ANS & Vega \\
7422.04978 & 17.066 & 0.061 & 0.94586 & ANS & Vega \\
7422.09426 & 17.516 & 0.022 & 0.98104 & Andicam I & Vega \\
7422.09605 & 17.846 & 0.012 & 0.98246 & Andicam B & Vega \\
7422.09780 & 17.769 & 0.022 & 0.98384 & Andicam V & Vega \\
7422.09905 & 17.513 & 0.022 & 0.98483 & Andicam R & Vega \\
7423.18165 & 17.034 & 0.018 & 0.84110 & Andicam I & Vega \\
7423.18340 & 17.343 & 0.010 & 0.84248 & Andicam B & Vega \\
7423.18515 & 17.324 & 0.019 & 0.84386 & Andicam V & Vega \\
7423.18640 & 17.249 & 0.019 & 0.84485 & Andicam R & Vega \\
7424.04142 & -1.000 & -1.000 & 0.52112 & ANS & Vega \\
7424.04142 & 16.656 & 0.020 & 0.52112 & ANS & Vega \\
7424.04142 & 16.956 & 0.022 & 0.52112 & ANS & Vega \\
7424.11649 & 16.952 & 0.016 & 0.58049 & Andicam I & Vega \\
\hline
\end{tabular}
\end{table}
\begin{table}[]
\contcaption{}
\begin{tabular}{@{}lrrrrrrrrrl}
\hline
Date$^a$ & mag & mag&Orbital$^c$& instrument  &phot.&  \\
(MJD)    &     & error    &phase      & and filter   & system &  \\
\hline
7424.11824 & 17.066 & 0.008 & 0.58188 & Andicam B & Vega \\
7424.11999 & 17.043 & 0.015 & 0.58326 & Andicam V & Vega \\
7424.12124 & 16.963 & 0.016 & 0.58425 & Andicam R & Vega \\
7424.12285 & 16.833 & 0.006 & 0.58552 & OGLE I & Vega \\
7425.12780 & 16.926 & 0.017 & 0.38037 & Andicam I & Vega \\
7425.12955 & 17.170 & 0.008 & 0.38175 & Andicam B & Vega \\
7425.13134 & 17.149 & 0.016 & 0.38317 & Andicam V & Vega \\
7425.13255 & 17.025 & 0.018 & 0.38412 & Andicam R & Vega \\
7425.16908 & 16.887 & 0.006 & 0.41302 & OGLE I & Vega \\
7426.15560 & 17.134 & 0.021 & 0.19329 & Andicam I & Vega \\
7426.15735 & 17.224 & 0.009 & 0.19467 & Andicam B & Vega \\
7426.15914 & 17.192 & 0.017 & 0.19609 & Andicam V & Vega \\
7426.16035 & 17.093 & 0.019 & 0.19705 & Andicam R & Vega \\
7426.16622 & 16.974 & 0.008 & 0.20169 & OGLE I & Vega \\
7427.14781 & 17.542 & 0.009 & 0.97806 & OGLE I & Vega \\
7428.10171 & 17.178 & 0.018 & 0.73254 & Andicam I & Vega \\
7428.10346 & 17.414 & 0.009 & 0.73392 & Andicam B & Vega \\
7428.10521 & 17.378 & 0.017 & 0.73531 & Andicam V & Vega \\
7428.10646 & 17.250 & 0.019 & 0.73629 & Andicam R & Vega \\
7428.15238 & 17.202 & 0.007 & 0.77261 & OGLE I & Vega \\
7429.11102 & 17.266 & 0.020 & 0.53083 & Andicam I & Vega \\
7429.11277 & 17.449 & 0.010 & 0.53221 & Andicam B & Vega \\
7429.11452 & 17.458 & 0.018 & 0.53360 & Andicam V & Vega \\
7429.11577 & 17.358 & 0.020 & 0.53459 & Andicam R & Vega \\
7429.16175 & 17.156 & 0.006 & 0.57096 & OGLE I & Vega \\
7431.11144 & 17.391 & 0.007 & 0.11303 & OGLE I & Vega \\
7431.44867 & 17.194 & 0.036 & 0.37976 & UVOT UVW1 & AB \\
7431.51464 & 17.184 & 0.034 & 0.43194 & UVOT UVW1 & AB \\
7431.84659 & 17.529 & 0.040 & 0.69448 & UVOT UVW1 & AB \\
7431.91256 & 17.453 & 0.037 & 0.74666 & UVOT UVW1 & AB \\
7432.12850 & 17.775 & 0.010 & 0.91746 & OGLE I & Vega \\
7432.37992 & 17.470 & 0.036 & 0.11632 & UVOT UVW1 & AB \\
7432.77229 & 17.344 & 0.028 & 0.42665 & UVOT UVW1 & AB \\
7433.30354 & 17.717 & 0.034 & 0.84683 & UVOT UVW1 & AB \\
7433.83410 & 17.547 & 0.058 & 0.26647 & UVOT UVW1 & AB \\
7434.10890 & 17.560 & 0.009 & 0.48382 & OGLE I & Vega \\
7434.37299 & 17.592 & 0.034 & 0.69270 & UVOT UVW1 & AB \\
7434.43132 & 17.762 & 0.055 & 0.73883 & UVOT UVW1 & AB \\
7435.14184 & 17.698 & 0.010 & 0.30081 & OGLE I & Vega \\
7435.22646 & 17.722 & 0.034 & 0.36774 & UVOT UVW1 & AB \\
7435.69174 & 17.962 & 0.034 & 0.73574 & UVOT UVW1 & AB \\
7436.29105 & 18.058 & 0.034 & 0.20976 & UVOT UVW1 & AB \\
7436.82578 & 17.938 & 0.035 & 0.63269 & UVOT UVW1 & AB \\
7437.11450 & 18.227 & 0.015 & 0.86105 & OGLE I & Vega \\
7437.13404 & 18.188 & 0.030 & 0.87650 & Andicam I & Vega \\
7437.13579 & 18.591 & 0.042 & 0.87789 & Andicam B & Vega \\
7437.13759 & 18.557 & 0.051 & 0.87931 & Andicam V & Vega \\
7437.13879 & 18.450 & 0.043 & 0.88026 & Andicam R & Vega \\
7437.74939 & 17.938 & 0.040 & 0.36321 & UVOT UVW1 & AB \\
7438.08501 & 18.074 & 0.033 & 0.62866 & Andicam I & Vega \\
7438.08680 & 18.284 & 0.036 & 0.63008 & Andicam B & Vega \\
7438.08855 & 18.322 & 0.047 & 0.63146 & Andicam V & Vega \\
7438.08980 & 18.169 & 0.040 & 0.63245 & Andicam R & Vega \\
7438.28134 & 18.310 & 0.054 & 0.78394 & UVOT UVW1 & AB \\
7438.75426 & 18.231 & 0.048 & 0.15799 & UVOT UVW1 & AB \\
7439.03049 & 18.436 & 0.034 & 0.37647 & Andicam I & Vega \\
7439.03224 & 18.409 & 0.026 & 0.37785 & Andicam B & Vega \\
7439.03399 & 18.750 & 0.054 & 0.37924 & Andicam V & Vega \\
7439.03524 & 18.134 & 0.049 & 0.38023 & Andicam R & Vega \\
7439.07718 & 18.113 & 0.046 & 0.41340 & UVOT UVW1 & AB \\
7439.14844 & 18.109 & 0.016 & 0.46976 & OGLE I & Vega \\
\hline
\end{tabular}
\end{table}
\begin{table}[]
\contcaption{}
\begin{tabular}{@{}lrrrrrrrrrl}
\hline
Date$^a$ & mag & mag&Orbital$^c$& instrument  &phot.&  \\
(MJD)    &     & error    &phase      & and filter   & system &  \\
\hline
7439.61190 & 18.517 & 0.053 & 0.83633 & UVOT UVW1 & AB \\
7440.08900 & 18.407 & 0.046 & 0.21368 & Andicam I & Vega \\
7440.09075 & 18.555 & 0.058 & 0.21507 & Andicam B & Vega \\
7440.09250 & 18.558 & 0.063 & 0.21645 & Andicam V & Vega \\
7440.09375 & 18.287 & 0.059 & 0.21744 & Andicam R & Vega \\
7440.27579 & 18.234 & 0.055 & 0.36142 & UVOT UVW1 & AB \\
7440.66955 & 18.477 & 0.114 & 0.67285 & UVOT UVW1 & AB \\
7441.02710 & 18.946 & 0.048 & 0.95565 & Andicam I & Vega \\
7441.02885 & 19.208 & 0.078 & 0.95704 & Andicam B & Vega \\
7441.03064 & 19.505 & 0.090 & 0.95845 & Andicam V & Vega \\
7441.03193 & 19.299 & 0.068 & 0.95948 & Andicam R & Vega \\
7441.10293 & 18.912 & 0.035 & 0.01563 & OGLE I & Vega \\
7441.26816 & 18.558 & 0.054 & 0.14632 & UVOT UVW1 & AB \\
7441.73413 & 18.353 & 0.051 & 0.51487 & UVOT UVW1 & AB \\
7442.04882 & 18.085 & 0.036 & 0.76377 & Andicam I & Vega \\
7442.05057 & 18.749 & 0.051 & 0.76516 & Andicam B & Vega \\
7442.05236 & 18.418 & 0.058 & 0.76657 & Andicam V & Vega \\
7442.05357 & 18.637 & 0.050 & 0.76753 & Andicam R & Vega \\
7442.39803 & 18.703 & 0.060 & 0.03997 & UVOT UVW1 & AB \\
7442.59942 & 18.553 & 0.061 & 0.19925 & UVOT UVW1 & AB \\
7443.06192 & 18.506 & 0.051 & 0.56506 & UVOT UVW1 & AB \\
7443.09063 & 18.443 & 0.025 & 0.58777 & Andicam I & Vega \\
7443.09354 & 18.653 & 0.030 & 0.59007 & Andicam B & Vega \\
7443.09646 & 18.646 & 0.026 & 0.59238 & Andicam V & Vega \\
7443.09942 & 18.420 & 0.024 & 0.59472 & Andicam R & Vega \\
7443.11736 & 18.283 & 0.015 & 0.60891 & OGLE I & Vega \\
7443.12006 & 18.260 & 0.015 & 0.61105 & OGLE I & Vega \\
7443.65915 & 18.707 & 0.052 & 0.03743 & UVOT UVW1 & AB \\
7444.05776 & 18.544 & 0.048 & 0.35271 & UVOT UVW1 & AB \\
7444.65498 & 18.769 & 0.058 & 0.82507 & UVOT UVW1 & AB \\
7445.10716 & 18.410 & 0.021 & 0.18271 & OGLE I & Vega \\
7445.58555 & 18.727 & 0.058 & 0.56108 & UVOT UVW1 & AB \\
7446.02288 & 18.822 & 0.022 & 0.90698 & Andicam I & Vega \\
7446.02580 & 19.104 & 0.013 & 0.90929 & Andicam B & Vega \\
7446.02876 & 19.037 & 0.013 & 0.91163 & Andicam V & Vega \\
7446.03167 & 18.932 & 0.014 & 0.91394 & Andicam R & Vega \\
7446.58208 & 18.709 & 0.053 & 0.34928 & UVOT UVW1 & AB \\
7447.04343 & 18.760 & 0.021 & 0.71418 & Andicam I & Vega \\
7447.04635 & 19.181 & 0.013 & 0.71649 & Andicam B & Vega \\
7447.04931 & 18.919 & 0.012 & 0.71883 & Andicam V & Vega \\
7447.05226 & 18.733 & 0.013 & 0.72116 & Andicam R & Vega \\
7447.13940 & 18.686 & 0.018 & 0.79008 & OGLE I & Vega \\
7448.02695 & 18.666 & 0.020 & 0.49207 & Andicam I & Vega \\
7448.02987 & 19.180 & 0.012 & 0.49438 & Andicam B & Vega \\
7448.03278 & 18.816 & 0.012 & 0.49668 & Andicam V & Vega \\
7448.03574 & 18.747 & 0.013 & 0.49902 & Andicam R & Vega \\
7448.57723 & 18.717 & 0.063 & 0.92731 & UVOT UVW1 & AB \\
7449.01826 & 18.687 & 0.026 & 0.27613 & Andicam I & Vega \\
7449.02118 & 19.233 & 0.014 & 0.27844 & Andicam B & Vega \\
7449.02413 & 18.792 & 0.014 & 0.28077 & Andicam V & Vega \\
7449.02705 & 18.831 & 0.016 & 0.28308 & Andicam R & Vega \\
7449.10640 & 18.584 & 0.049 & 0.34585 & UVOT UVW1 & AB \\
7449.12580 & 18.459 & 0.022 & 0.36119 & OGLE I & Vega \\
7449.57168 & 18.710 & 0.054 & 0.71385 & UVOT UVW1 & AB \\
7450.01960 & 18.771 & 0.023 & 0.06813 & Andicam I & Vega \\
7450.02252 & 19.008 & 0.012 & 0.07044 & Andicam B & Vega \\
7450.02547 & 18.906 & 0.013 & 0.07277 & Andicam V & Vega \\
7450.02839 & 18.871 & 0.015 & 0.07508 & Andicam R & Vega \\
7450.03835 & 18.803 & 0.062 & 0.08296 & UVOT UVW1 & AB \\
7450.56752 & 18.696 & 0.053 & 0.50150 & UVOT UVW1 & AB \\
7451.10166 & 18.873 & 0.023 & 0.92396 & Andicam I & Vega \\
\hline
\end{tabular}
\end{table}
\begin{table}[]
\contcaption{}
\begin{tabular}{@{}lrrrrrrrrrl}
\hline
Date$^a$ & mag & mag&Orbital$^c$& instrument  &phot.&  \\
(MJD)    &     & error    &phase      & and filter   & system &  \\
\hline
7451.10462 & 19.208 & 0.014 & 0.92630 & Andicam B & Vega \\
7451.10758 & 19.187 & 0.013 & 0.92865 & Andicam V & Vega \\
7451.11049 & 19.006 & 0.014 & 0.93095 & Andicam R & Vega \\
7451.16683 & 19.312 & 0.076 & 0.97551 & UVOT UVW1 & AB \\
7452.02922 & 18.656 & 0.024 & 0.65760 & Andicam I & Vega \\
7452.03213 & 18.898 & 0.012 & 0.65990 & Andicam B & Vega \\
7452.03509 & 18.904 & 0.012 & 0.66224 & Andicam V & Vega \\
7452.03801 & 18.889 & 0.015 & 0.66455 & Andicam R & Vega \\
7453.04974 & 18.816 & 0.021 & 0.46476 & Andicam I & Vega \\
7453.05269 & 19.243 & 0.013 & 0.46709 & Andicam B & Vega \\
7453.05561 & 18.963 & 0.013 & 0.46940 & Andicam V & Vega \\
7453.05853 & 18.963 & 0.015 & 0.47171 & Andicam R & Vega \\
7453.10025 & 18.718 & 0.019 & 0.50471 & OGLE I & Vega \\
7453.49185 & 19.068 & 0.110 & 0.81444 & UVOT UVM2 & AB \\
7453.49532 & 18.899 & 0.109 & 0.81719 & UVOT UVW1 & AB \\
7453.49671 & 18.253 & 0.104 & 0.81829 & UVOT U & AB \\
7453.49810 & 19.300 & 0.086 & 0.81939 & UVOT UVW2 & AB \\
7454.03666 & 19.264 & 0.088 & 0.24535 & Andicam B & Vega \\
7454.03958 & 19.125 & 0.079 & 0.24766 & Andicam V & Vega \\
7454.04249 & 18.981 & 0.029 & 0.24997 & Andicam R & Vega \\
7455.88284 & 18.955 & 0.156 & 0.70556 & UVOT UVM2 & AB \\
7455.88423 & 18.822 & 0.160 & 0.70666 & UVOT UVW1 & AB \\
7455.88492 & 18.074 & 0.153 & 0.70721 & UVOT U & AB \\
7455.88562 & 19.110 & 0.112 & 0.70775 & UVOT UVW2 & AB \\
7455.94881 & 18.834 & 0.098 & 0.75774 & UVOT UVM2 & AB \\
7455.95159 & 18.715 & 0.100 & 0.75993 & UVOT UVW1 & AB \\
7455.95367 & 18.417 & 0.122 & 0.76158 & UVOT U & AB \\
7455.95506 & 19.098 & 0.076 & 0.76268 & UVOT UVW2 & AB \\
7456.01465 & 18.848 & 0.025 & 0.80981 & Andicam I & Vega \\
7456.01757 & 19.081 & 0.014 & 0.81212 & Andicam B & Vega \\
7456.02053 & 19.213 & 0.015 & 0.81446 & Andicam V & Vega \\
7456.02348 & 18.934 & 0.016 & 0.81679 & Andicam R & Vega \\
7457.01450 & 18.757 & 0.023 & 0.60062 & Andicam I & Vega \\
7457.01746 & 19.018 & 0.013 & 0.60296 & Andicam B & Vega \\
7457.02042 & 18.965 & 0.013 & 0.60530 & Andicam V & Vega \\
7457.02338 & 18.739 & 0.013 & 0.60765 & Andicam R & Vega \\
7457.11303 & 18.722 & 0.022 & 0.67856 & OGLE I & Vega \\
7457.41619 & 18.944 & 0.089 & 0.91833 & UVOT UVM2 & AB \\
7457.42035 & 18.860 & 0.089 & 0.92163 & UVOT UVW1 & AB \\
7457.42313 & 18.644 & 0.103 & 0.92382 & UVOT U & AB \\
7457.42452 & 19.066 & 0.063 & 0.92492 & UVOT UVW2 & AB \\
\hline
\end{tabular}
\begin{tabular}{@{}lrrrrrrrrrl}
$^a$ The full set of collected photometric and spectroscopic data on N\ LMC 1968 will be submitted to the IAU astronomical data centers. \\ 
$^b$ time of exposure; mid-time if possible.  \\
$^c$ from eq.~\ref{eq1}. \\
\\
\end{tabular}

\end{table}

\bsp
\label{lastpage}
\end{document}